\documentclass[vci-paper]{vciart}
\usepackage{amssymb}

\begin{document}

\begin{frontmatter}

\title{New Developments and Future Perspectives of Gaseous Detectors}

% else, use optional labels to link authors explicitly to addresses,
% as shown below:
\author[A,B]{Maxim Titov} 
\address[A]{CEA/SACLAY, DAPNIA, 91191 Gif-sur-Yvette, France}
\address[B]{Albert-Ludwigs University of Freiburg, Physics Institute, Freiburg, Germany}

\begin{abstract}

 Gaseous detectors are fundamental components of all present 
and planned high energy physics experiments.
 Over the past decade two representatives (GEM, Micromegas) of the
Micro-Pattern Gas Detector (MPGD) concept have become
increasingly important; the high radiation resistance and
excellent spatial and time resolution 
make them an invaluable tool
to confront future detector challenges at the next generation of colliders.
 Novel structures where GEM and Micromegas are directly coupled to the 
CMOS multi-pixel readout represent an exciting field
and allow to reconstruct fine-granularity, two-dimensional images of physics events.
Originally developed for the high energy physics,
MPGD applications have expanded to 
astrophysics, neutrino physics, neutron detection and medical imaging.

\end{abstract}

\end{frontmatter}

\section{Introduction}

 The compelling scientific goals of future high energy physics experiments
are a driving factor in the development of advanced detector technologies.
 With many fundamental issues within the experimental reach at the
Large Hadron Collider (LHC), 
a large $R\&D$ effort was devoted to 
the optimization of existing devices and the development of
innovative concepts for radiation detection.
 All four major LHC experiments (ATLAS, CMS, ALICE and LHCb)
will extensively use gas detectors,
depending on the need for trigger and bunch crossing tagging (GEM, Resistive Plate
Chambers, Thin Gap Chambers), high-rate tracking and particle identification (Straw-Type
Detectors, Time Projection Chamber, Drift Tubes, Cathode Strip Chambers) 
and single photon detection (Multi-Wire Proportional Chamber with CsI photocathodes).
 In future colliders, gaseous detectors are planned to be used
at the ten times higher luminosity upgrade of LHC (the so-called superLHC or SLHC)
or at the foreseen International Linear Collider (ILC).
 The related $R\&D$ challenges come on one hand 
from the development of radiation-hard detector concepts and
high granularity readout
and on the other hand from the need for ultra-high-precision tracking
with minimum material budget.

 Within the broad family of different gas detector technologies,
two innovative MPGD concepts, GEM and Micromegas,
have now reached maturity and 
play a prominent role in modern HEP experiments.
 Their attractive properties such as 
excellent spatial resolution, fast signal response, low radiation length and
high rate capability combined with radiation hardness, motivate
their use at the future SLHC and ILC colliders.
 There are many common issues to be addressed 
at the next generation of $e^+e^-$ and $pp$ machines, 
necessitating more fundamental $R\&D$
in the field of micro-pattern devices, which would be
beneficial for both communities.
 This paper reviews recent advances of GEM and Micromegas
detectors, with a focus on design principles, performance, 
and operational experience,
and discusses the most promising directions in future developments and
applications.

\section{Gaseous detectors at the LHC: System Aspects of Modern Experiments}

The development of modern LHC detectors presented formidable challenges, which are 
at the same time of great technological, engineering and organizational complexity.
 Typical time-scales have stretched over 20 years
starting from the concept, progressing through intensive $R\& D$, 
design and prototyping, mass production, installation 
and finally system integration and commissioning. 
 These detectors are not just bigger versions of currently running experiments;
several new issues had to be addressed during their construction:
\begin{itemize}
\item Radiation hardness 
at extremely high radiation levels;
\item Intrinsic performance of innovative detector technologies 
integrated into large systems 
at the physics frontier;
\item A coherent and system oriented approach from the $R\&D$ phase to
commissioning, including many-year-long strong partnership with industry; 
\item Sub-detector systems become so large and complex that their construction
has to be shared around the world.
\end{itemize}
 The LHC gaseous detectors will not be reviewed in this paper;
they are covered in a number of dedicated articles~\cite{virdee,froidevaux}.
 Although achieving the intrinsic performance in the lab captures
most of experimenter's attention, 
the complexity of the overall system approach
remains the area where most systems experience major setbacks. 
 Understanding of all system aspects should be a focus 
of vigorous $R\&D$ to develop common solutions for upgraded and new experiments.
 The commissioning of the LHC experiments
is about to reach a successful completion over the coming year.
 Using the ATLAS Muon Spectrometer as an example of a modern gaseous detector,
the main
challenges of achieving the intrinsic performance in a large system
using standard wire chamber technology are discussed.

\subsection{ATLAS Muon Detector: Modern Large-Volume Spectrometer}

 The ATLAS Muon Spectrometer has been designed for standalone tracking with a 
momentum resolution of 2.5$\%$ for transverse momenta up to 250~GeV (limited by 
multiple scattering) and better than 10$\%$ up to 1~TeV (limited by the muon chamber 
measurement accuracy)~\cite{atlas_spectrometer}.
 A total of 370~000 Monitored Drift Tubes (MDT), arranged in 1200 chambers and 
covering an active area of more than 5500~m$^{2}$, 
have been built at 13 construction sites worldwide over a period of 5 years~\cite{dubbert}. 
 The basic detection element of the ATLAS MDT is an aluminum 
tube of 3~cm diameter and 500~$\mu m$ wall thickness, with a 50~$\mu m$ diameter
central gold-plated W-Re wire. 
 The detector is operated with a low longitudinal diffusion 
$Ar/CO_2$ (93:7) mixture at 3~bar absolute pressure. 
 Despite various improvements, position-sensitive detectors 
based on wire structures are 
limited by basic diffusion processes and space charge 
effects in the gas to localization accuracies around 
100~$\mu m$.
 The presence of slow-moving ions from electron avalanches 
generates a positive space charge in the drift tube, 
which modifies the electric field 
and leads to an uncertainty in the space-to-drift-time relation.
  This results in a degradation of the single-tube resolution with increasing
irradiation rate, especially for large impact radii (see Fig.~\ref{atlasmdt}a).
 Close to the sense wire the space charge effects lower the electric field
causing a gain drop.

\setlength{\unitlength}{1mm}
\begin{figure}[bth]
 \begin{picture}(50,50)
 \put(7.0,-2.0){\includegraphics{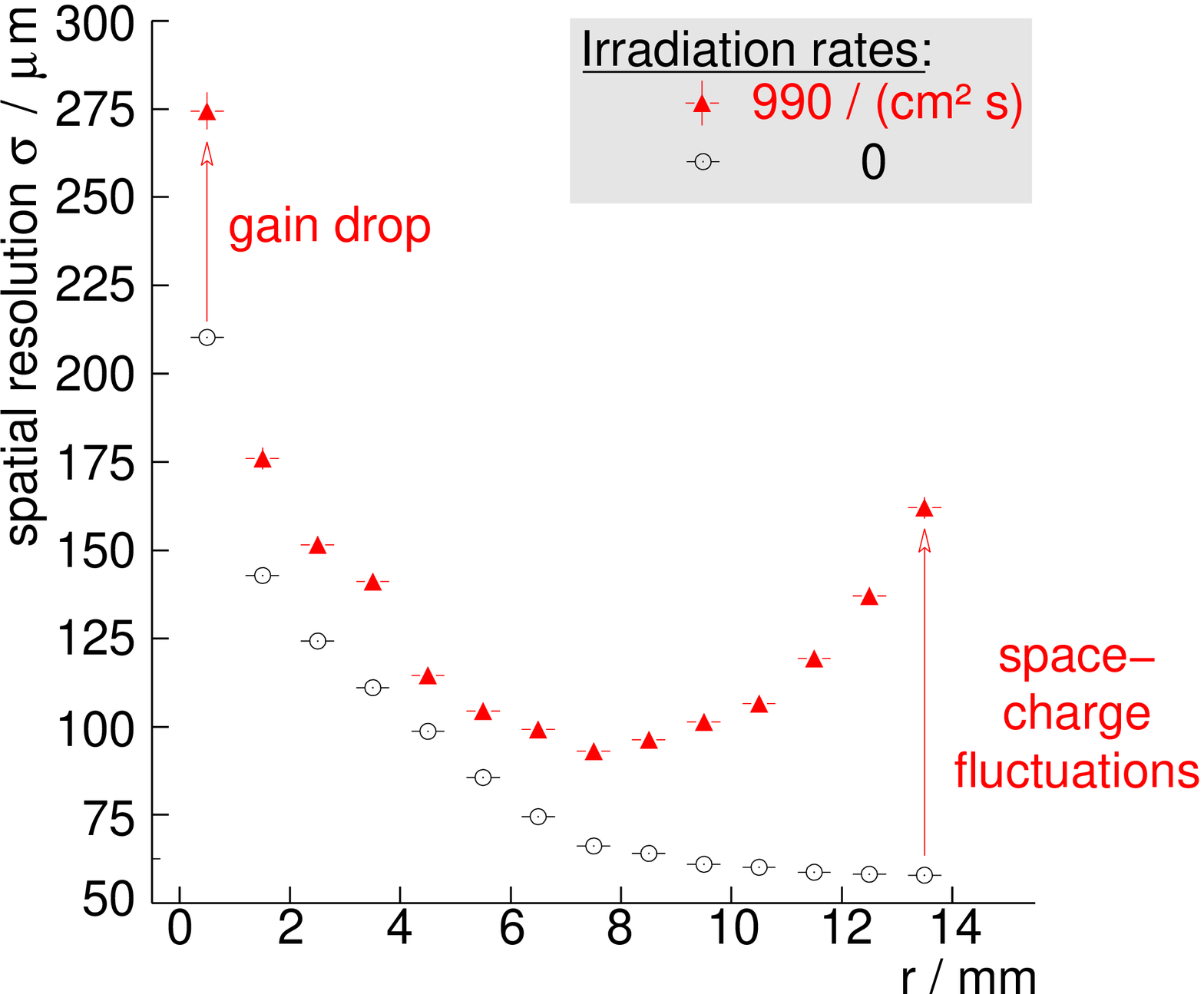}}
 \put(79.0,-3.0){\includegraphics{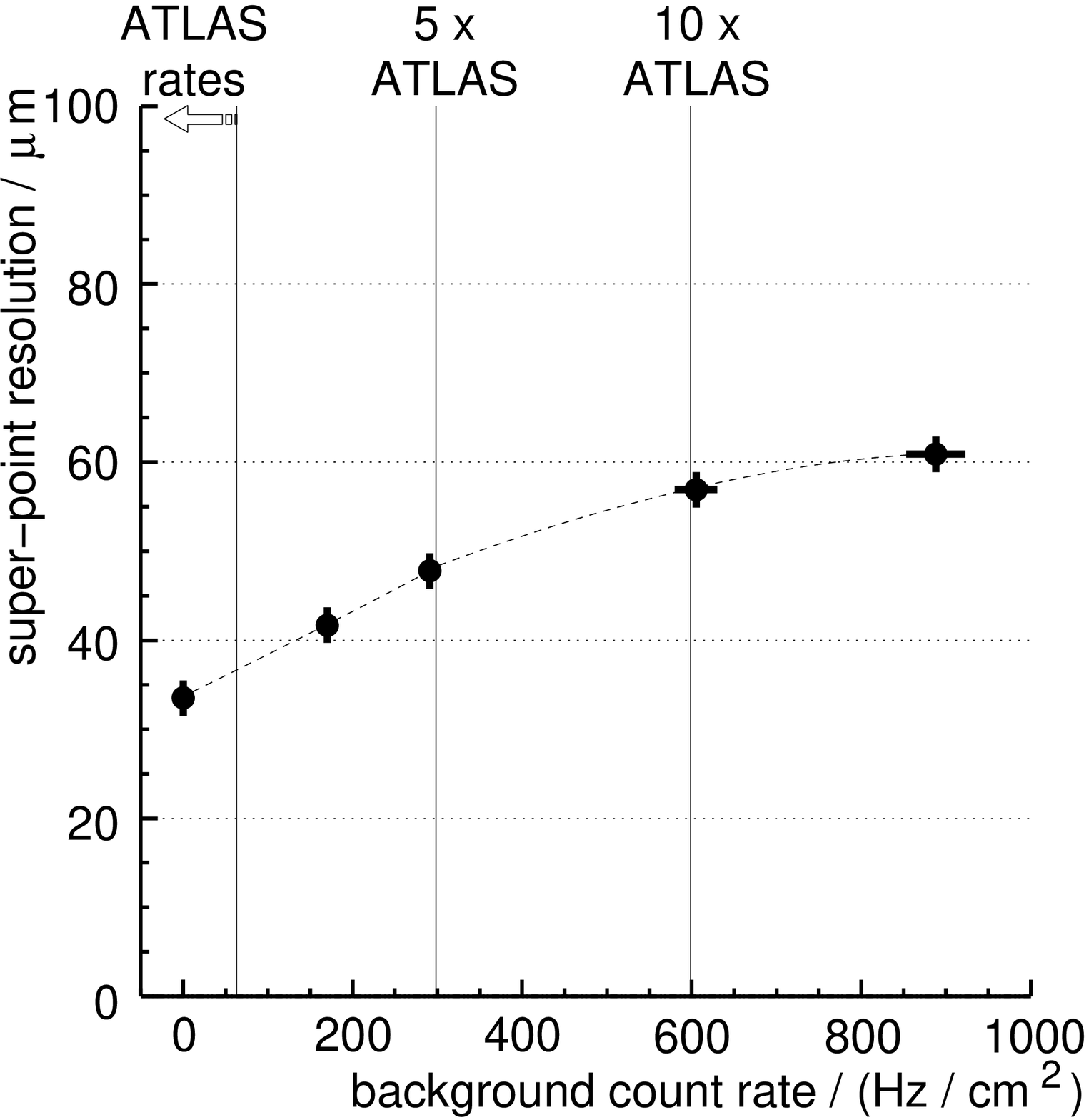}}
 \put(-1.0,48.0){ a) }
 \put(70.0,48.0){ b) }
 \end{picture}
\caption{ a) Single tube resolution 
as a function of the muon impact radius, with and without irradiation.
b) Position resolution of the MDT chamber 
as a function of the background particle flux in the ATLAS Muon Spectrometer.}
\label{atlasmdt}
\end{figure}

The momentum resolution goals in the ATLAS Muon Spectrometer
imply an overall precision of $\sim$50~$\mu m$
on the track points, given the available bending power.
 To improve the resolution of a chamber beyond the single wire limit,
the MDT chambers are constructed from
2$\times$3 monolayers of drift tubes,
glued on either side of a rigid support structure.
 The high accuracy of the standalone measurement 
puts stringent demands on the mechanical precision 
of the sense wire location (20~$\mu m$).
 In addition, an internal alignment system with 12000 optical sensors 
has to monitor continuously the deformations and movements of the 
precision chambers with a relative accuracy of 30~$\mu m$.
 The space-drift-time relation 
for a single tube has 
to be known with an accuracy of 20 $\mu m$. 
 Finally, the inhomogeneity of the magnetic field throughout the whole detector volume
must be measured and monitored to an accuracy of approximately 20~G.
Recent studies have demonstrated that it is feasible to reach the required system performance, 
as shown in Fig.~\ref{atlasmdt}b.
 A spatial resolution of $\sigma<$~60~$\mu m$ per MDT chamber 
was measured
at particle fluxes of up to 600~$Hz/cm^2$, which corresponds to 10~times
the highest background rate expected in the ATLAS Muon Spectrometer.

\section{Micro-Pattern Gaseous Detectors}

 Modern photo-lithographic technology has enabled a series of inventions of novel 
MPGD concepts: Micro-Strip Gas Chamber (MSGC), GEM, Micromegas and many others~\cite{mpgd_workshop},
revolutionizing cell size limits for many gas detector applications.
  The MSGC, a concept invented in 1988 by A. Oed~\cite{msgc}, was the first 
of the microstructure gas detectors. 
 Consisting of a set of tiny metal strips 
laid on a thin insulating substrate, and alternatively connected as anodes and cathodes,
the MSGC turned out to be easily damaged by discharges
induced by heavily ionizing particles and destroying the fragile electrode structure~\cite{bagaturia}.
 The more powerful GEM and Micromegas concepts
fulfill the needs of high-luminosity colliders with increased reliability in harsh
radiation environments.
 By using smaller feature size compared to classical gas counters,
these detectors offer 
intrinsic high rate capability
(fine pitch and fast collection of positive ions)~\cite{benloch,giomataris4},
excellent spatial resolution ($\sim$30~$\mu m$)~\cite{gem_beamtest,derre1}, 
and single-photoelectron time resolution in the nanosecond range~\cite{breskin_1e,derre_csi}.

\setlength{\unitlength}{1mm}
\begin{figure}[bth]
 \begin{picture}(40,40)
 \put(5.0,-5.0){\includegraphics{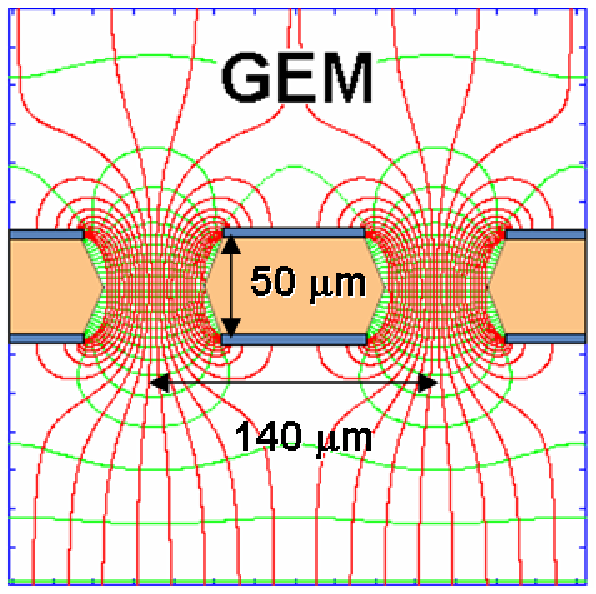}}
 \put(70.0,-8.0){\includegraphics{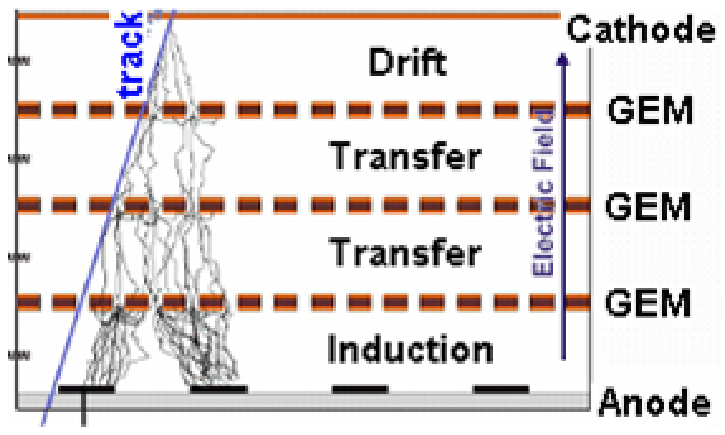}}
 \put(-1.0,37.0){ a) }
 \put(65.0,37.0){ b) }
 \end{picture}
\caption{ a) Schematics and electric field map of the GEM amplification cell.
b) Schematical drawing of the triple-GEM detector.}
\label{GEM_Micromegas_sketch1}
\end{figure}

 Introduced in 1996 by F. Sauli~\cite{sauli1997}, a GEM consists of a set of holes, arranged in a
hexagonal pattern (typically 70~$\mu m$ diameter at 140~$\mu m$ pitch),
chemically etched through copper-kapton-copper thin-foil composite.
 Application of a potential difference between the two sides of the GEM generates the
field map shown in Fig.~\ref{GEM_Micromegas_sketch1}a: electrons
released by the ionization in the gas drift into the holes and multiply in the high 
electric field (50-70~kV/cm).
 Sharing the avalanche multiplication among several cascaded 
electrodes (see Fig.~\ref{GEM_Micromegas_sketch1}b)
allows to operate triple-GEM detectors at overall gains above $10^4$
in the presence of highly ionizing particles
while eliminating the risk of hazardous discharges ($< 10^{-12}$ per hadron). 
This is 
the major advantage of the GEM technology~\cite{gem_discharge1,gem_discharge2}.
 A unique property of the GEM detector is the complete decoupling of the amplification stage (GEM)
and the readout electrode (PCB), which operates at unity gain and serves only as a charge collector.
 This offers some freedom in the optimization of the anode readout structure, which can be made of pads 
or strips of arbitrary pattern~\cite{nima478}.
GEMs can also be easily bent to form 
cylindrically curved ultra-light detectors, 
as preferred for inner tracker applications~\cite{ropelewski,bencivenni}.
 Controlled etching of GEM foils (decreasing the thickness of the copper layer from 5 to 1 $\mu m$)
allows to reduce the material budget in triple GEMs to 1.5$\times$10$^{-3} X_0$,
which is about one half of a 300-$\mu m$-thick Si-microstrip detector~\cite{bondar1}.
 Recently, several companies (Tech Etch and 3M) have started to develop
an industrial version of GEMs~\cite{becker}. Their mechanical and physical properties
are found to be similar to the foils produced at the CERN workshop~\cite{azmoun}.

\setlength{\unitlength}{1mm}
\begin{figure}[bth]
 \begin{picture}(40,40)
 \put(2.0,-7.0){\includegraphics{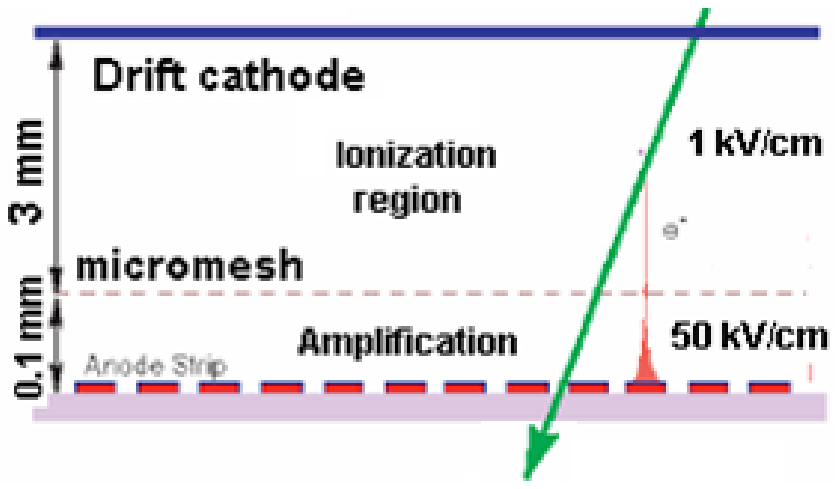}}
 \put(73.0,-5.0){\includegraphics{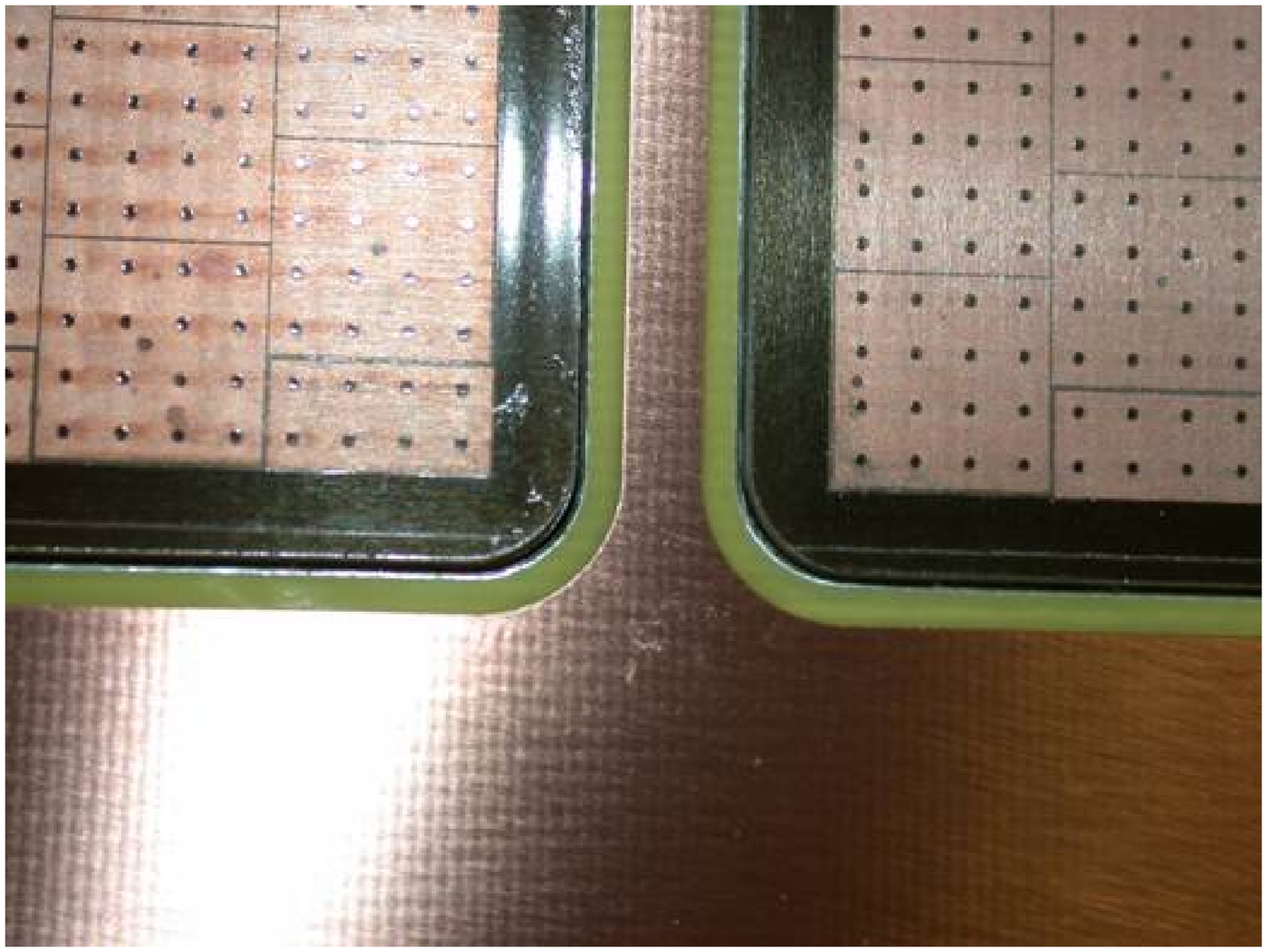}}
 \put(-1.0,35.0){ a) }
 \put(65.0,35.0){ b) }
 \end{picture}
\caption{ a) Schematical drawing of the Micromegas detector.
b) Photograph of the ``Bulk'' Micromegas detectors. Pillars of 400~$\mu m$ diameter
every 2~mm are visible.}
\label{GEM_Micromegas_sketch2}
\end{figure}

 Introduced in 1996 by I. Giomataris~\cite{giomataris1996}, 
a micro-mesh gaseous structure (Micromegas) 
is a double stage parallel plate avalanche counter (see Fig.~\ref{GEM_Micromegas_sketch2}a).
It consists of a few mm conversion region (electric field $\sim$~1~kV/cm) and a 
narrow multiplication gap (25-150~$\mu m$, 50-70~kV/cm), located between 
a thin metal grid (micromesh) and the readout electrode
(strips/pads of conductor printed on an insulator board).
 To preserve a distance between the anode and the grid mesh,
spacers made of insulating material are used.
 The small amplification gap is a key element in Micromegas operation and gives rise 
to excellent spatial resolution: 12~$\mu m$ accuracy (limited by the pitch of micromesh)
is achieved for MIPs with a strip pitch of 100~$\mu m$ and low diffusion
$CF_4/iC_4H_{10}$ (80:20) mixture~\cite{derre1}.
 Micromegas exploits the saturation characteristics of the Townsend coefficient 
at high fields to achieve a reduced dependence of the gas gain on gap variations,
leading to very good energy resolution ($\sim$12~$\%$ FWHM at 6~keV)~\cite{delpart1}.

 A big step in the direction of the industrial manufacturing of large-size detectors
is the development of the ``Bulk'' Micromegas technology~\cite{giomataris2}.
 The basic idea is to built the whole detector in a single process:
the anode plane with copper strips, a photo-imageable polyamid film
and the woven mesh are laminated together at high temperature forming a single object.
 At the end, the micromesh is sandwiched between 2 layers of insulating material,
which is removed after UV exposure and chemical development.
 Several large ``Bulk'' Micromegas (27*26 cm$^2$), as prototypes for T2K/TPC,
have been produced (see Fig.~\ref{GEM_Micromegas_sketch2}b) and 
successfully tested in the HARP TPC field cage at CERN inside a magnetic field~\cite{bouchez,sarrat}.
 This technique has been recently extended 
to build even larger-size detectors, up to 65$\times$55~$cm^2$,
in a single piece~\cite{giomataris3}.

\subsection{High-Rate Tracking and Triggering}

 COMPASS is a first high-luminosity experiment at CERN which pioneered the use of 
GEM and Micromegas detectors for high-rate particle tracking, 
reaching 25~$kHz/mm^2$ in the near-beam area.
 Both technologies have achieved a tracking efficiency of close to 100~$\%$,
a spatial resolution of the order of 70 - 100~$\mu m$ 
and a time resolution of $\sim$~10~ns~\cite{gemcompass},~\cite{micromegascompass}.
 The excellent performance and radiation hardness of 22 large-size triple-GEM (30*30 cm$^2$)
and 12 Micromegas (40*40 cm$^2$) detectors after several years of successful operation
has demonstrated the large-scale reliability and robustness of the MPGD concept.
 No degradation of performance is observed in COMPASS detectors
after an accumulated charge of a few milliCoulombs/$mm^2$,
corresponding to an equivalent flux of $\sim10^{11}$ MIPs/$mm^2$.
 For the COMPASS physics program in 2007, a set of triple-GEM trackers 
with pixel readout (1$\times$1~$mm^2$) in the central region and 2D strip readout in
the periphery is being built~\cite{ketzer_vienna}.

 The intrinsic time resolution of GEM and Micromegas detectors is determined 
by the time distribution of the primary ionization clusters and the signal
amplitude fluctuations, both in the ionization and multiplication processes.
A time resolution of $\sim 10~ns$ (RMS) measured in the COMPASS GEM detectors  
is limited by the moderate electron drift velocity in $Ar/CO_2$ 
and by the readout electronics.
 Using fast $CF_4$-based mixtures, a time resolution of about 5~ns (RMS)~\cite{barouch}
and an efficiency of 96~$\%$ in a 20~ns time window can be achieved~\cite{lhcb}, 
adequate to resolve two bunch crossings at the high luminosity colliders.
 GEMs have entered the LHC program; 
they will be used for triggering in the
LHCb Muon System~\cite{lhcb} and 
in the TOTEM Telescopes~\cite{ropelewski,totem}.

\subsection{Time Projection Chamber Readout}

 The Time Projection Chamber (TPC) concept, invented in 1976~\cite{nygren}, 
has been the prime choice for large tracking systems in $e^+ e^-$ colliders (PEP-4, ALEPH, DELPHI)
and proved its unique resolving power 
in heavy-ion collisions (NA35, NA49 and STAR). 
 It is an ideal tracker with minimal material budget before the calorimeters
for high-multiplicity topologies occurring at low rates.
 A TPC consists of a large gas volume, with a uniform electric field applied between 
the central electrode and a grid at the opposite side.
 The ionization trails produced by charged particles drift towards the readout end-plate
where a 2D image of tracks is reconstructed;
the third coordinate is measured using the drift time information.
 The large number of 3D space points makes for a robust and efficient tracking system,
even for tracks within the densely collimated jets at the future ILC, and
offers the capability to measure particle momenta and to perform 
particle identification through dE/dx measurements.
 A conventional readout structure, based on MWPC and pads, is a benchmark
for the ``most modern'' ALICE TPC, designed to cope with extreme 
instantaneous particle densities produced in heavy-ion collisions 
at the LHC.
 This detector incorporates innovative and state of the art technologies,
from the mechanical structures to the readout electronics and data processing 
chain~\cite{antonczyk,frankenfeld}.
 To limit distortions of the ALICE TPC intrinsic spatial resolution
($\sigma_{r\phi}\sim$1000~$\mu m$ for 250~cm drift length), 
the temperature gradient in an 88~$m^3$ gas volume space
must not exceed 0.1$^{\circ} C$~\cite{meyer}.

\setlength{\unitlength}{1mm}
\begin{figure}[bth]
 \begin{picture}(53,53)
 \put(0.0,-8.0){\includegraphics{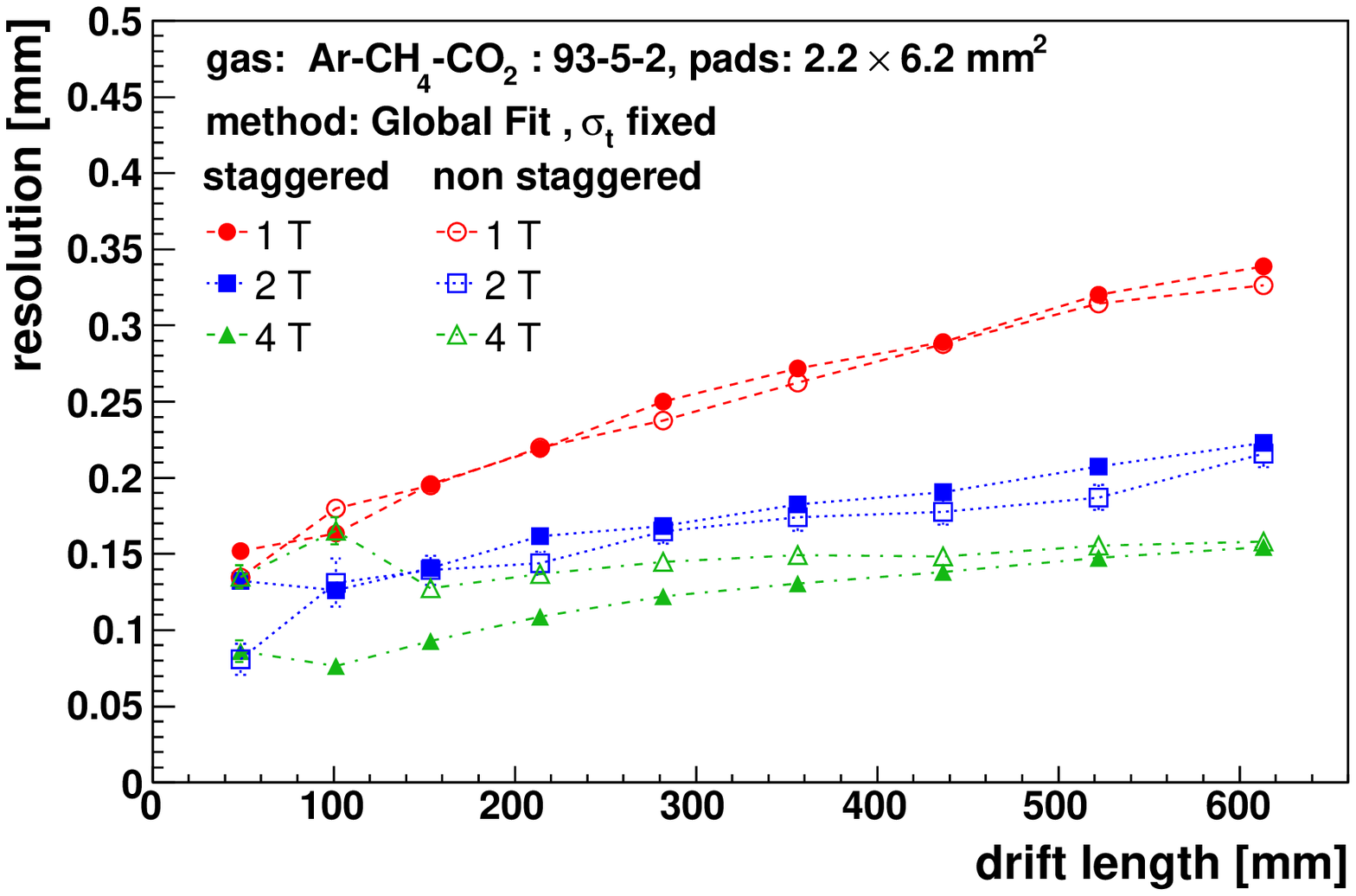}}
 \put(73.0,57.0){\includegraphics{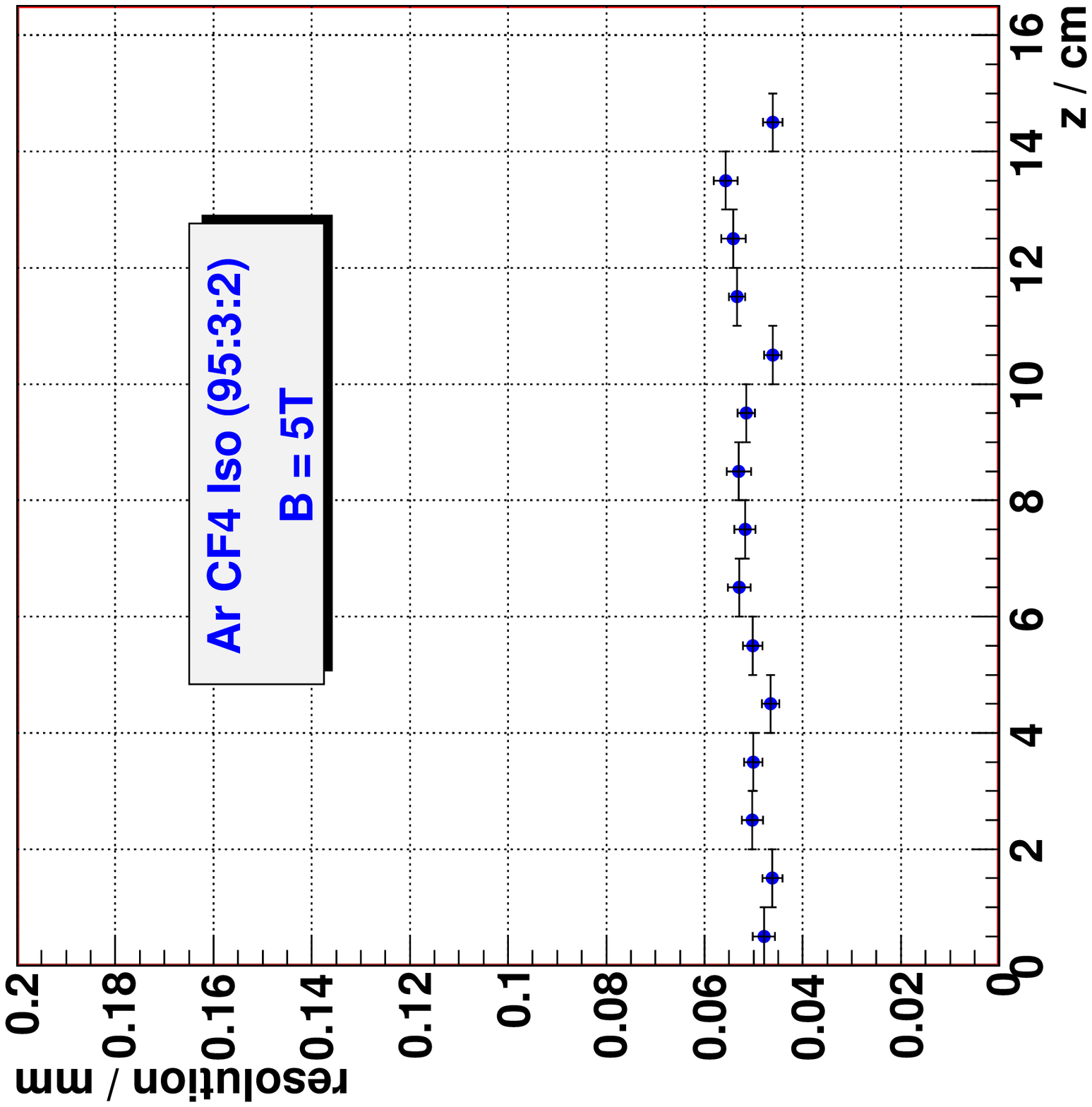}}
 \put(-1.0,48.0){ a) }
 \put(70.0,48.0){ b) }
 \end{picture}
\caption{ Transverse spatial resolution as a function of drift distance for 2$\times$6~$mm^2$ pads: 
a) GEM-TPC with the $Ar/CH_4/CO_2$(93:5:3) mixture at 1~T, 2~T and 4~T fields;
b) Micromegas-TPC, using charge dispersive readout, with $Ar/CF_4/iC_4H_{10}$(95:3:2) 
mixture in 5~T field.}
\label{ilctpc_resolution}
\end{figure}

$R\&D$ for a high-precision TPC is ongoing in the international ILC-TPC
collaboration~\cite{ilctpc}.
 The main topics are the construction of a low material-budget field cage and the 
development of gas amplification end-plates using GEM or Micromegas.
 In the context of the particle flow concept at the ILC, the requirements for charged
particle tracking are mainly high efficiency and double track resolution.
 In addition, a momentum resolution of $\sigma (1/p_t) \approx 10^{-4}/GeV $ in the TPC
alone has to be achieved to reconstruct Higgs mass in a model independent way
(exploiting the recoil mass technique) in Higgs-strahlung events
($e^+ e^- \rightarrow ZH \rightarrow llX$) and
to measure the end-point momentum in supersymmetric slepton decays
($\tilde{l} \tilde{l} \rightarrow \mathrm{\tilde \chi^0_1} \mathrm{\tilde \chi^0_1} ll$).
 Therefore, at least 200 space points with an average transverse resolution of 100~$\mu m$
have to be measured 
over a maximum drift length of 250~cm in a 4~T field,
and a multi-track separation of 1~mm has to be reached.
 These requirements are beyond the limits of MWPC but can be 
fulfilled with GEM or Micromegas readout, 
which offer a number of advantages:
negligible E$\times$B track distortion effects,
narrow pad response function (PRF)
and intrinsic ion feedback suppression.

 The principle of the ILC-TPC MPGD concept
has been successfully validated over the last years.
 The single point resolution of $\sim$100$~\mu m$ 
has been achieved with a GEM-TPC prototype after 60~cm of drift in a 4~T field
(see Fig.~\ref{ilctpc_resolution}a).
 The spatial resolution for low magnetic fields shows the expected 
dependence on the drift length, which is caused by diffusion.
 For high fields the GEM-TPC resolution is dominated by the PRF~\cite{janssen}.
 Recent studies with a Micromegas TPC using a charge-dispersive readout technique have
demonstrated an excellent single point
resolution of $50~\mu m$ over the 15~cm drift length in a 5~T field, as shown in
 Fig.~\ref{ilctpc_resolution}b~\cite{attie}.
 In this case a high-surface-resistivity thin film is laminated 
to the anode 
forming a distributed two-dimensional RC network with 
respect to the readout plane~\cite{dixit1}.
 The arriving avalanche charge at the anode disperses with the RC system
time constant (determined by the anode surface resistivity and capacitance per unit area).
 This avoids degradation of the point resolution
due to single-pad hits at short drift distances.
 A fractional ion back-flow into the drift volume down to 0.2~$\%$
is measured both with GEM and Micromegas, 
which in combination with modern low-noise electronics
might allow to avoid the use of a gating grid.
 An attractive approach to use a GEM foil for gating, if needed,
has been proposed in~\cite{sauli_ionfeedback}.
However, the effect of electron transmission losses on the spatial resolution
(at low GEM voltages used for gating) has still to be experimentally measured.
 Other issues to be addressed include
optimization of single-point and double-track resolution in presence of background,
demonstration of large-system performance in a 4~T field with control
of systematics, and end-plate design for minimal material.

 The future MPGD-TPC developments are not limited to the $R \& D$ for the linear 
collider: in fact, they will be used in a variety of applications.
 Employing the ``Bulk'' technology, 72 large Micromegas (34$\times$36~$cm^2$)
will be built for the T2K/TPC detector to instrument an area of almost 10~$m^2$.
TPCs will eventually disappear from hadron colliders
because of the long memory time and proneness to space charge accumulation.
However, they will prevail at future lepton and heavy-ion colliders~\cite{oda}
where they constitute one of the most cost-effective central tracking systems.

\section{Pixel Readout for Micro-Pattern Gas Detectors}

 Advances in the micro-electronics industry and advanced PCB technology have been very important
for the development of modern gas detectors with increasingly smaller pitch size.
The fine granularity and high-rate capability of micro-pattern devices
can be fully exploited using a high-density pixel readout with a size 
corresponding to the intrinsic width of the detected avalanche charge. 
 However, for a pixel pitch of the order of 100~$\mu m$, technological constraints
severely limit the maximum number of channels that can be brought to the external 
front-end electronics. 
 An elegant solution is to use a CMOS pixel chip 
assembled directly below the GEM or Micromegas amplification structure
and serving as an integrated charge collecting anode.
  With this arrangement avalanche electrons are collected on the top
metal layer of the CMOS ASIC; every input pixel is then directly connected
to the amplification, digitization and sparsification circuits 
integrated in the underlying active layers of the CMOS technology.
 Using this approach, gas detectors can reach the level of integration typical
of solid-state pixel devices.

 Particle detectors are designed to achieve the sensitivity required to study physics 
processes of interest. 
 The multi-pixel anode readout of micro-pattern gas detectors allows a true 2D image
reconstruction and opens novel detection opportunities in:
\begin{itemize}
\item Astronomical $X$-ray polarimetry (2-10~keV energy range);
\item Position sensitive single-electron detection;
\item Time Projection Chamber readout;
\item High-rate particle tracking;
\item Advanced Compton Telescopes (0.4-50~MeV energy range);
\item Low energy nuclear recoil reconstruction (WIMP interactions).
\end{itemize}

 The advent of finely segmented MPGD with pixel read-out
could lead to the appearance of a highly efficient $X$-ray polarimeter in the 2-10~keV energy band, 
which would allow to measure simultaneously position- 
and energy-resolved linear polarisation~\cite{bellazzini1}.
 The real breakthrough 
was the development of an analog, low-noise and high
granularity ($50 \mu m$ pitch) multi-pixel ASIC~\cite{bellazzini2}-~\cite{bellazzini5} 
shown in Fig.~\ref{ASIC_photo}a.
So the initial direction and dynamics of photoelectron energy loss
in the gas can be accurately tracked before they are distorted by Coulomb scattering.

\setlength{\unitlength}{1mm}
\begin{figure}[bth]
 \begin{picture}(51,51)
 \put(3.0,-5.0){\includegraphics{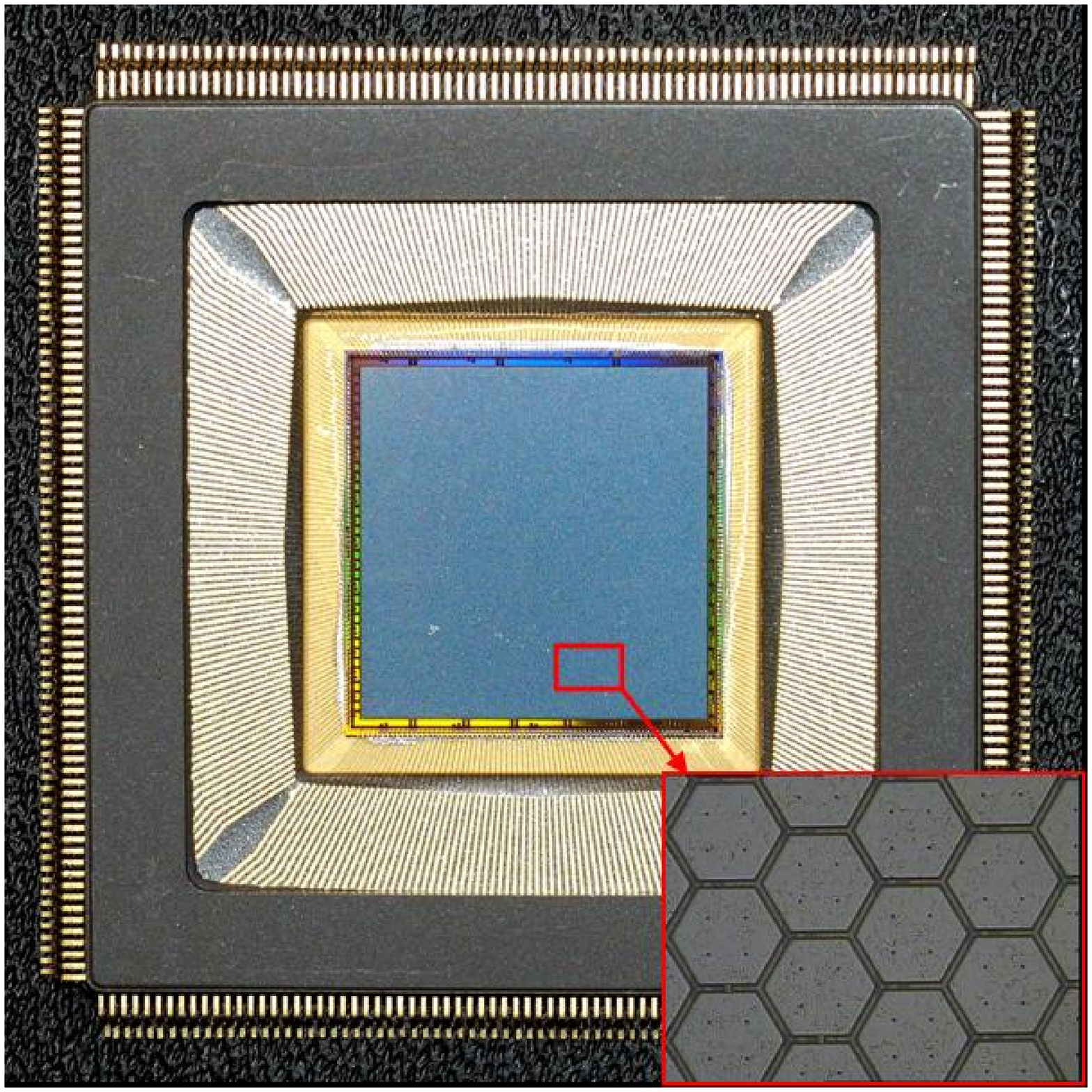}}
 \put(72.0,-5.0){\includegraphics{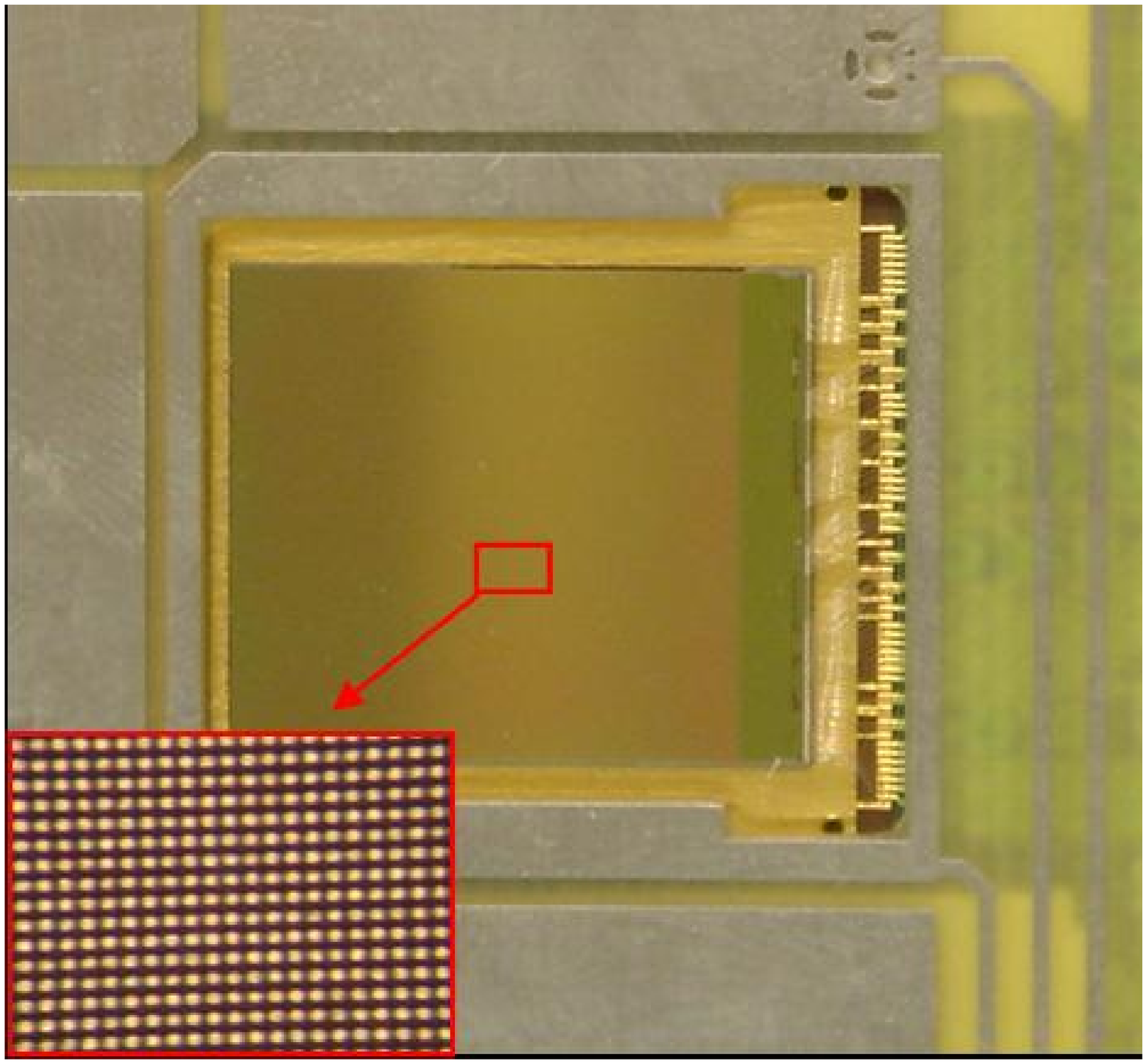}}
 \put(-1.0,48.0){ a) }
 \put(68.0,48.0){ b) }
 \end{picture}
\caption{ a) Photo of the analog CMOS ASIC with hexagonal pixels,
bonded to the ceramic package~\cite{bellazzini4}.
b) Photo of the Medipix2 chip~\cite{medipix2}; the 25~$\mu m$ wide conductive bump bond openings
used for electron collection are seen as a matrix of dots~\cite{freiburg1}.}
\label{ASIC_photo}
\end{figure}

 Following a similar approach, a binary multi-pixel CMOS chip (``Medipix2''), originally
developed for $X$-ray imaging~\cite{medipix2} has been shown to work with Micromegas and GEM
detectors~\cite{colas1}-\cite{freiburg1}. 
 Approximately 75~$\%$ of every pixel in the Medipix2 matrix are covered
with an insulating passivation layer. 
Hence, the avalanche electrons are collected
on the conductive bump-bonding pads,
% (octagonally shaped, 25 $\mu m$ wide)
exposed to the gas (see Fig.~\ref{ASIC_photo}b).
 Within the EUDET program~\cite{eudet}, a modification of the Medipix2 chip (``TimePix''~\cite{llopart})
which allows to measure the drift time information of primary electrons
has been designed, produced and already 
tested with GEM and Micromegas gas amplification systems.
 The Timepix chip uses an external clock up to 100~MHz as a time reference.
  Each pixel in the chip matrix can be programmed to record either
the electron arrival time with respect to an external shutter (``TIME'' mode) or 
the time-over-threshold (``TOT'' mode) information, providing a
pulse-height measurement.
 Developed as a potential readout for the ILC TPC to exploit the ultimate
spatial and double-track resolution, 
the TimePix chip can reconstruct the 3D-space points of individual electron clusters 
and thus count the number of ionization clusters per unit length for particle discrimination.
 Significant progress has also been made in the development of simulation tools and the 
comparison with data recorded with the Medipix2 and Timepix chips~\cite{hauschild}.
 Reading out large volume TPCs with highly segmented anode planes is also a key point 
for high-resolution track imagers proposed 
for an advanced Compton Telescope~\cite{nishimura,hattori} and for the 
detection of possible signatures of elastic interactions of WIMPs~\cite{sekiya,miuchi}.
 The primary advantage of a pixellated gas tracker is that the direction of the
Compton recoil electron or the low-energy nuclear recoil can be reconstructed 
far more accurately than in any other detection medium.

 One of the most exciting future applications of GEM and Micromegas devices with 
CMOS multi-pixel readout could be position sensitive single photon detection.
 The excellent spatial and time resolution, ambiguity-free reconstruction of multi-photon
events and non-negligible single-electron sensitivity
make them a suitable candidate for fast gas photo-multipliers.

 A key point that has to be solved to allow using CMOS pixel readout of MPGDs in high-energy physics
is the production of large area detectors.
 Recent progress in the development of edgeless silicon detectors~\cite{via} and the
possibility to bring power and I/O connections through the back of the
CMOS chip using the ``through-wafer vias'' technology~\cite{heijne,takahashi} 
may ultimately lead to the development of chips which are 4-side buttable.
 Properly integrated into large systems, the 
multi-pixel anode readout of micro-pattern
gas detectors may represent an invaluable tool
for the next generation of particle-physics experiments.
 However, a major $R\&D$ effort will be required in the future 
to fully exploit this potential.

\subsection{ GEM with a VLSI Pixel ASIC for $X$-Ray Astronomy}

 A GEM detector coupled to a VLSI analog pixel chip 
comprising a pixellated charge collecting electrode and readout electronics, 
can bring great improvement in sensitivity, at least 2 orders of magnitude,
compared to traditional $X$-ray polarimeters (based on Bragg crystals or
Compton scattering)~\cite{bellazzini1}.
The novel device allows to reconstruct individual
photoelectron tracks with a length as short as a
few hundred microns;
the total charge collected in the pixels
is proportional to the photon energy.
 The degree of $X$-ray polarization is computed  
from the distribution of reconstructed track angles since the
photoelectron is emitted mainly in the direction of the photon electric field.
 Three ASIC generations of increased complexity and size, reduced pitch and 
improved functionality
have recently been designed and built~\cite{bellazzini2}-\cite{bellazzini4}.
 The third ASIC version, realized in 0.18~$\mu m$ CMOS technology, 
includes a self-triggering capability and has 
105600 hexagonal pixels with 50~$\mu m$ pitch, 
corresponding to an active area of 15$\times$15~$mm^2$.
 A GEM coupled to such a CMOS pixel array 
is able to simultaneously produce high-resolution images ($50~\mu m$) and allow
moderate spectroscopy (15$\%$ FWHM at 6~keV) as well as fast timing (30~ns) 
in the 2-10~keV $X$-ray energy range.
At the focal plane of the large-area mirror of the XEUS telescope, 
a single 15$\times$15~$mm^2$ ASIC will cover a field of view of 5~arcmin, 
large enough to image extended objects like the Crab Nebula.
 Because of the high detector sensitivity, the polarization of Active Galactic Nuclei
down to the few-$\%$ level can be measured for 1 mCrab sources in one day~\cite{bellazzini4}.
 A sealed gas pixel detector for $X$-ray astronomy is currently under 
development~\cite{bellazzini5}.

\setlength{\unitlength}{1mm}
\begin{figure}[bth]
 \begin{picture}(51,51)
 \put(5.0,-3.0){\includegraphics{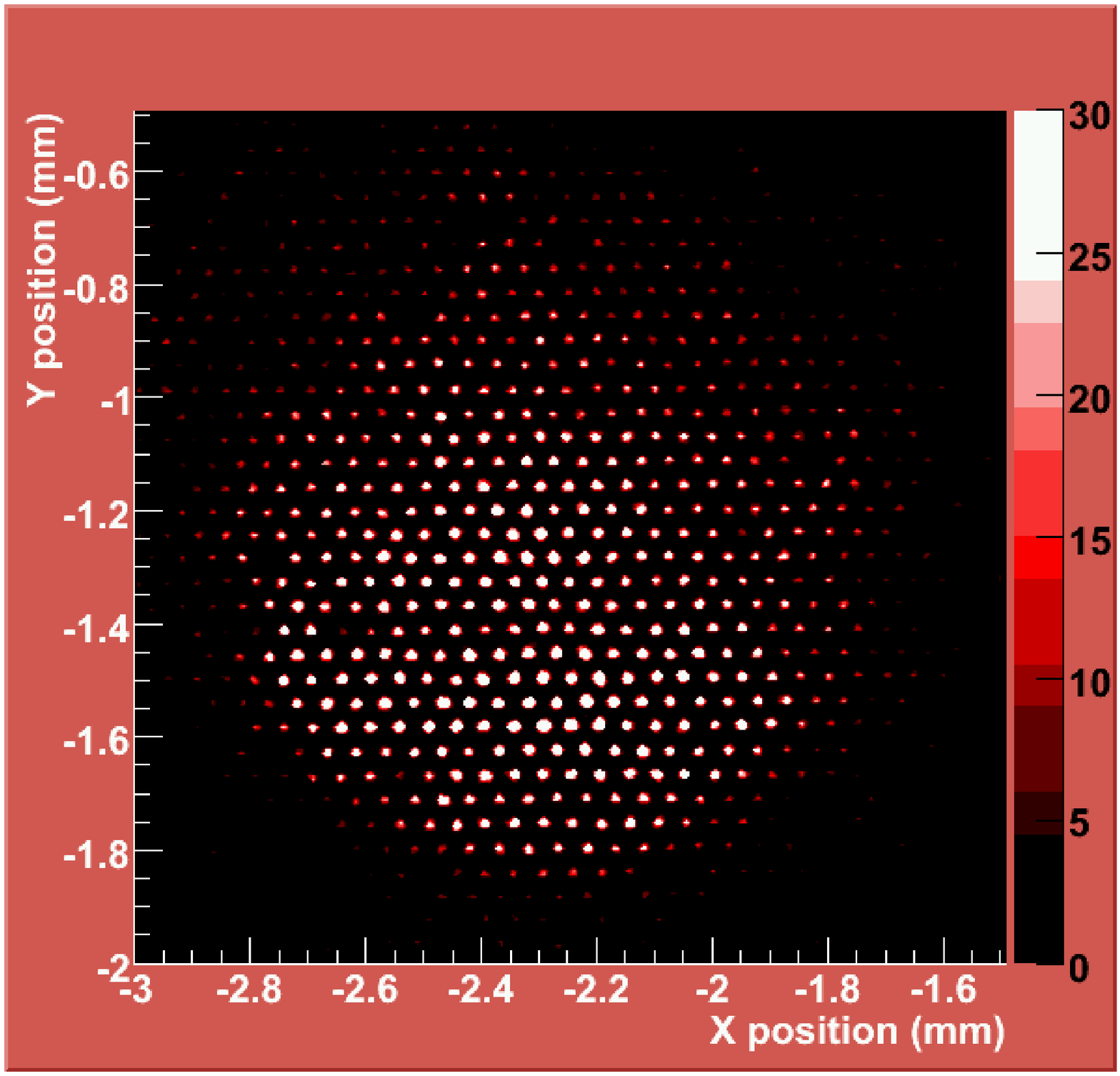}}
 \put(72.0,-7.0){\includegraphics{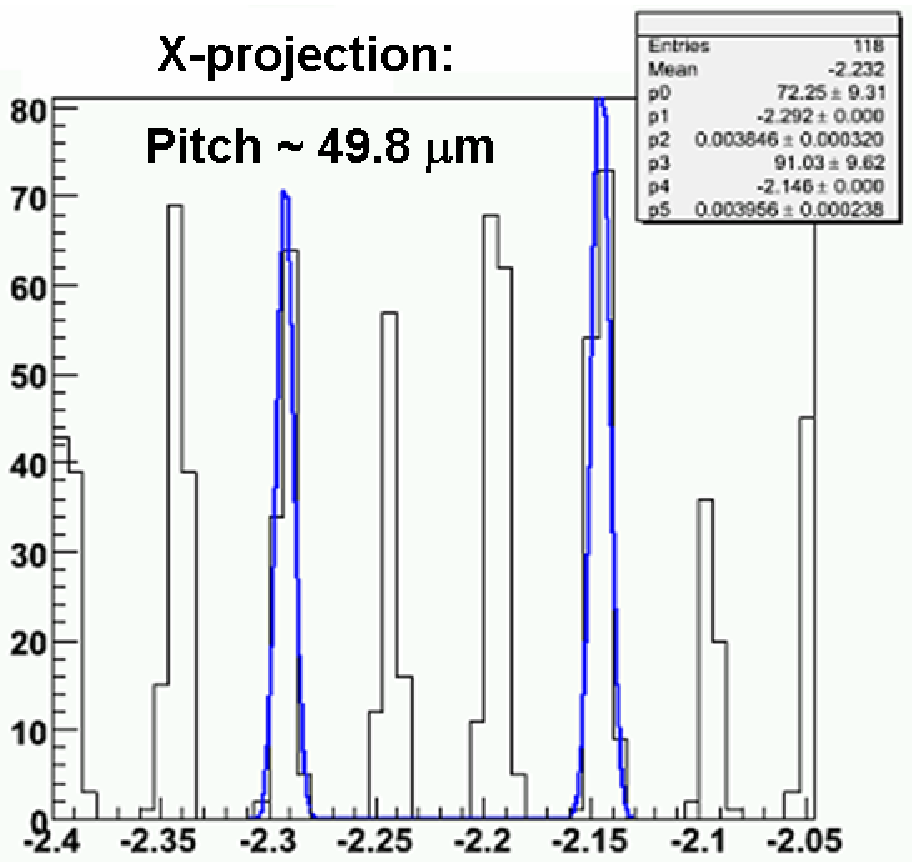}}
 \put(-1.0,48.0){ a) }
 \put(65.0,48.0){ b) }
 \end{picture}
\caption{ a) ``Self-portrait'' of the GEM amplification structure obtained with a
CMOS analog pixel chip.
b) A horizontal cut through the 2D image of barycenter positions 
(``centre of gravity'' of the single electron avalanche).
The peak-to-peak distance corresponds to the GEM pitch.
The gaussian fit yields $\sim$~4~$\mu m$ (rms) width~\cite{bellazzini_imaging}.}
\label{bellazzini_UVphoton}
\end{figure}

 Recently, 
a UV photo-detector based on a semitransparent CsI photocathode followed 
by a fine-pitch GEM foil matching the pitch of a pixel ASIC (50~$\mu m$)
has shown excellent imaging capabilities~\cite{bellazzini_imaging}.
 The photoelectron emitted from the CsI layer drifts into a 
single GEM hole and initiates an avalanche, which is then collected on the
pixel CMOS analog chip.
 Due to the high granularity and large $S/N$ of the read-out system,
the ``center of gravity'' of the single electron avalanche corresponds to 
the center of GEM hole.
 Accumulating thousands of such events produces the
``self-portrait'' of the GEM amplification structure
shown in Fig.~\ref{bellazzini_UVphoton}~(a).
The peaks, corresponding to GEM holes, are well resolved in the barycenter distribution in 
Fig.~\ref{bellazzini_UVphoton}~(b) allowing 
to achieve a superior single-electron avalanche reconstruction accuracy of 4~$\mu m$ (rms).
 Thanks to the very low pixel capacitance at the preamplifier input 
(noise $\sim 50 e^-$ ENC), the detector has significant sensitivity to a 
single primary electron even at a gas gain of a few thousand.
 The position resolution of the device is currently limited by the 50~$\mu m$ GEM pitch. 
The symmetric shape of a single-electron charge cloud at the readout plane
demonstrates that the
spatial resolution is not degraded by the avalanche spread inside the GEM
and is independent of the direction of entrance of the electron into the hole.

\subsection{ Micromegas Readout with Medipix2 / TimePix CMOS Chips}

 Initial 'proof-of-principal' studies using Micromegas foils equipped with the Medipix2
chip provided 2D images of minimum ionizing track clusters~\cite{colas1,campbel1}.
 The single (primary) electron efficiency was estimated to be around 90~$\%$;
the number of clusters in a $He/iC_4H_{10}$ (80:20) mixture was found to
agree within 15~$\%$ with simulation.
  The Timepix chip was also demonstrated to function perfectly with Micromegas;
3D track images from radioactive sources and cosmic rays have been observed
in the NIKHEF setup (see Fig.~\ref{Micromegas_pixel}a)~\cite{timmermans}.
 Prior to operation the Timepix chip was covered with a highly resistive 
4~$\mu m$ layer of amorphous silicon as protection against sparks 
and discharges.

\setlength{\unitlength}{1mm}
\begin{figure}[bth]
 \begin{picture}(50,50)
 \put(4.0,-5.0){\includegraphics{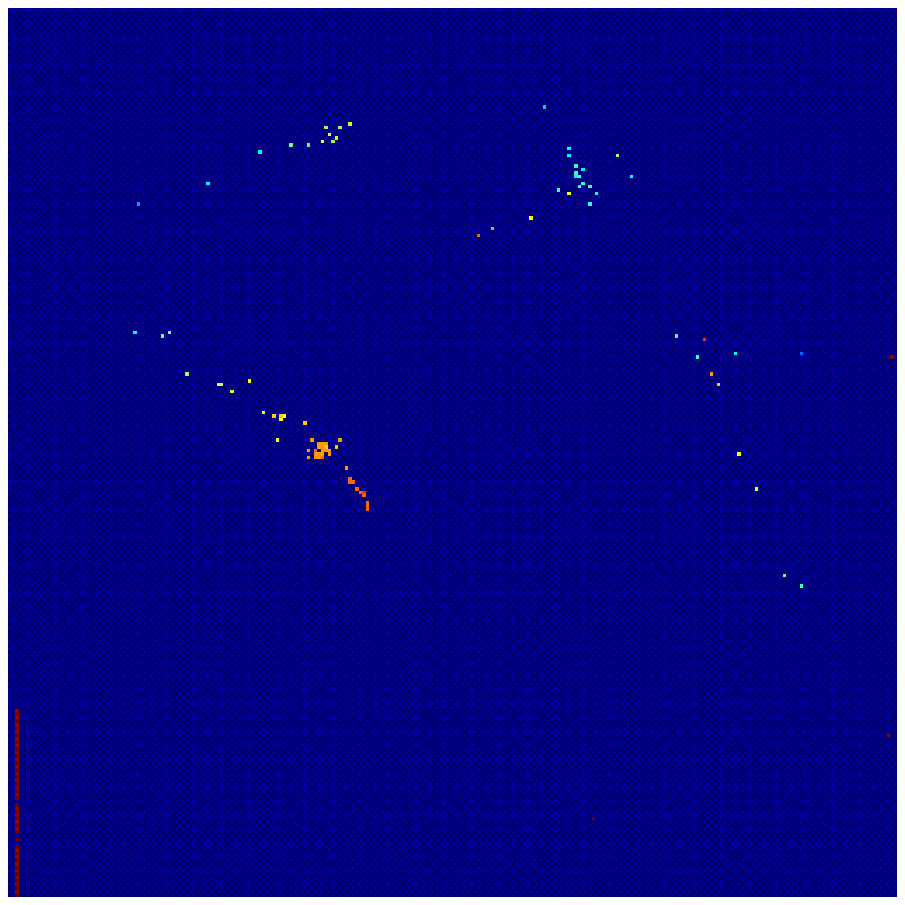}}
 \put(70.0,-5.0){\includegraphics{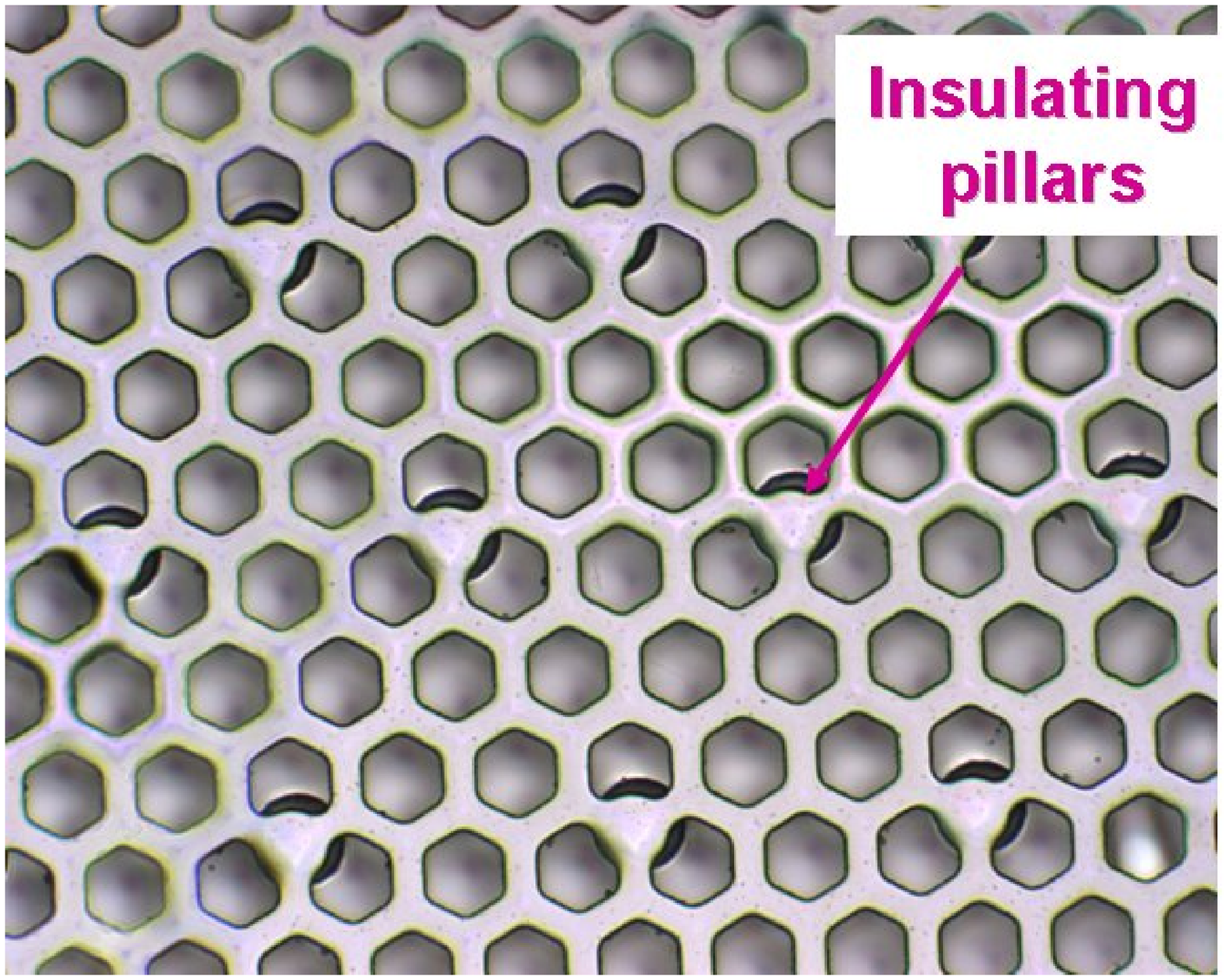}}
 \put(-1.0,45.0){ a) }
 \put(65.0,45.0){ b) }
 \end{picture}
\caption{ a) Cosmic rays recorded with a Micromegas/Timepix detector (``TOT'' mode); 
the color is a measure of the arrival time of the electrons~\cite{timmermans}.
b) Top view of InGrid structure with hexagonal holes (pitch 60~$\mu m$).
The insulating pillars (SU-8 epoxy) are centered between the grid holes.}
\label{Micromegas_pixel}
\end{figure}
 
 An attractive solution for the construction of MPGDs with pixel anode readout is the
integration of the Micromegas amplification and CMOS chip by means of the 'wafer
post-processing' technique~\cite{chefdeville}. 
 With this technology, the structure of a thin (1 $\mu m$) aluminum grid 
is fabricated on top of an array of insulating (SU-8) pillars of typically 50~$\mu m$
height standing above the CMOS chip and forming an integrated readout of the
gaseous detector (InGrid).
 The sub-$\mu m$ precision of the grid dimensions and avalanche gap size
results in a uniform gas gain; the grid hole size, pitch and pattern can be 
easily adapted to match the geometry of any pixel readout chip.
 The 'wafer post-processing' technology can also be used
if the readout CMOS matrix does not exactly match 
the required detector granularity.
 ``Through-wafer vias'' connections with variable re-routing lines allow to use 
detector elements with slightly smaller readout chips and 
space left over for external connections~\cite{heijne}.

\subsection{ Triple-GEM Readout with Medipix2 / Timepix CMOS Chips}

The triple-GEM detector with a Medipix2 chip has been initially 
studied in the Freiburg University with $^{55}$Fe $X$-Rays and $^{106}$Ru 
electrons~\cite{freiburg1}. 
  Stable operation at the gas gain of up to several $10^5$ has been achieved with
$Ar(He)/CO_2$ (70:30) mixtures.
  The device allows to perform moderate energy spectroscopy measurements
(20 $\%$ FWHM at 5.9 keV $X$-rays) using only digital readout and two
discriminator thresholds.
  A sample of $^{106}$Ru $\beta^-$-tracks was collected 
and the point resolution was evaluated using various methods and
taking into account the
multiple scattering contribution of electrons with a few MeV.
A spatial resolution of $\sim$50~$\mu m$ (averaged over tracks in the 6~mm drift gap) 
has been achieved, based on the binary centroid determination of the charge clouds.

\setlength{\unitlength}{1mm}
\begin{figure}[bth]
 \begin{picture}(53,53)
 \put(0.0,-7.0){\includegraphics{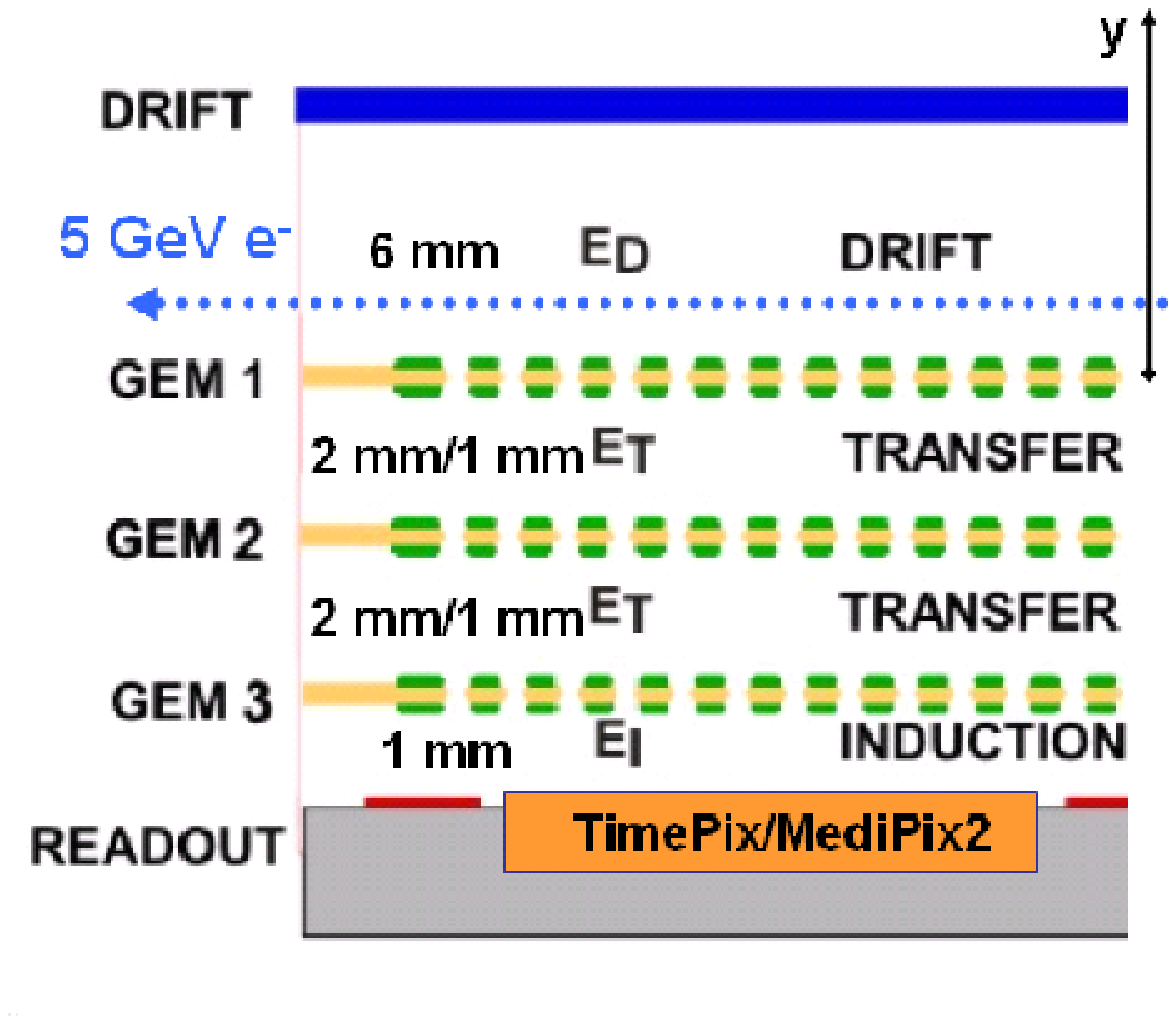}}
 \put(45.0,-2.0){\includegraphics{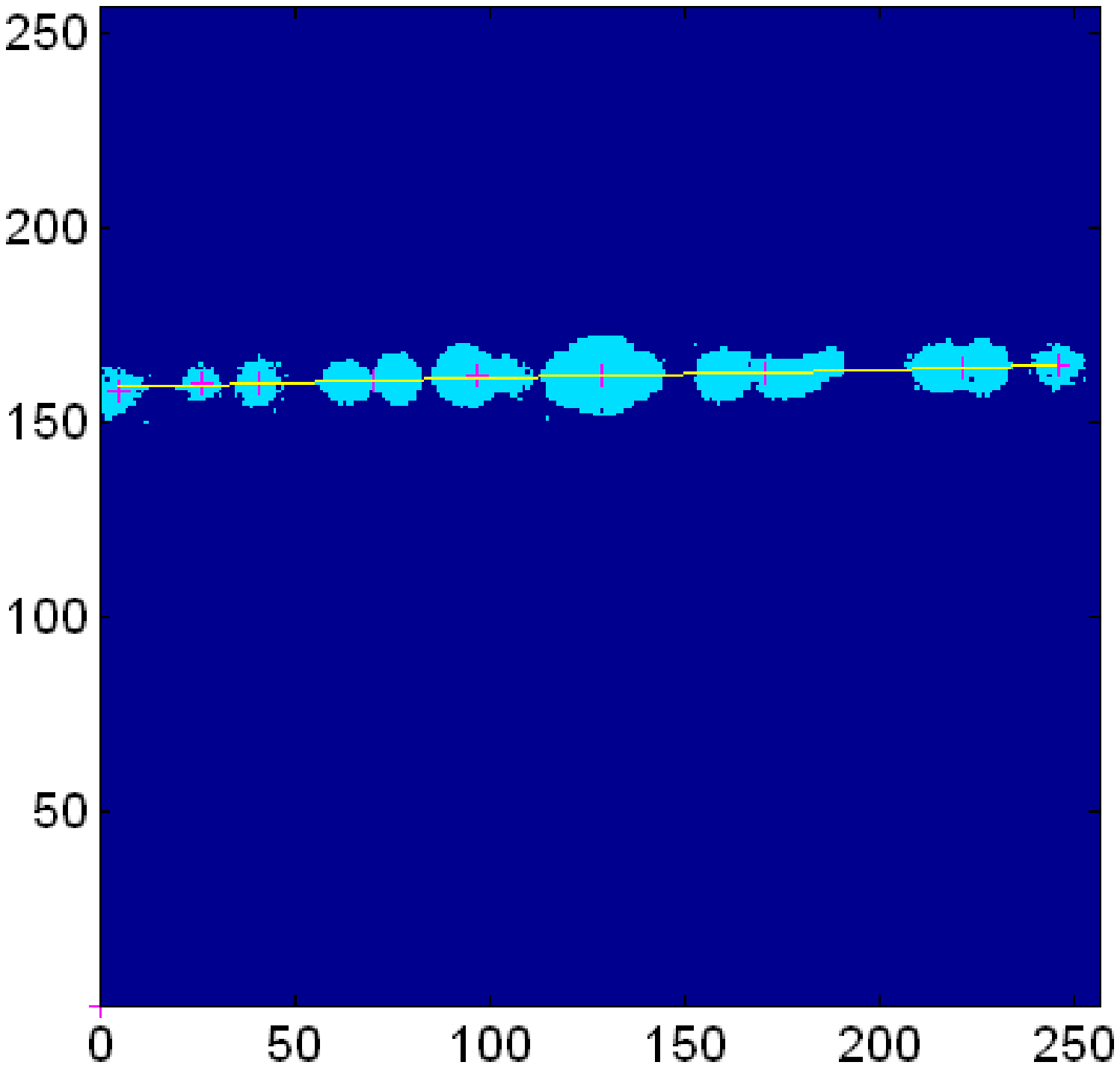}}
 \put(88.0,-23.0){\includegraphics{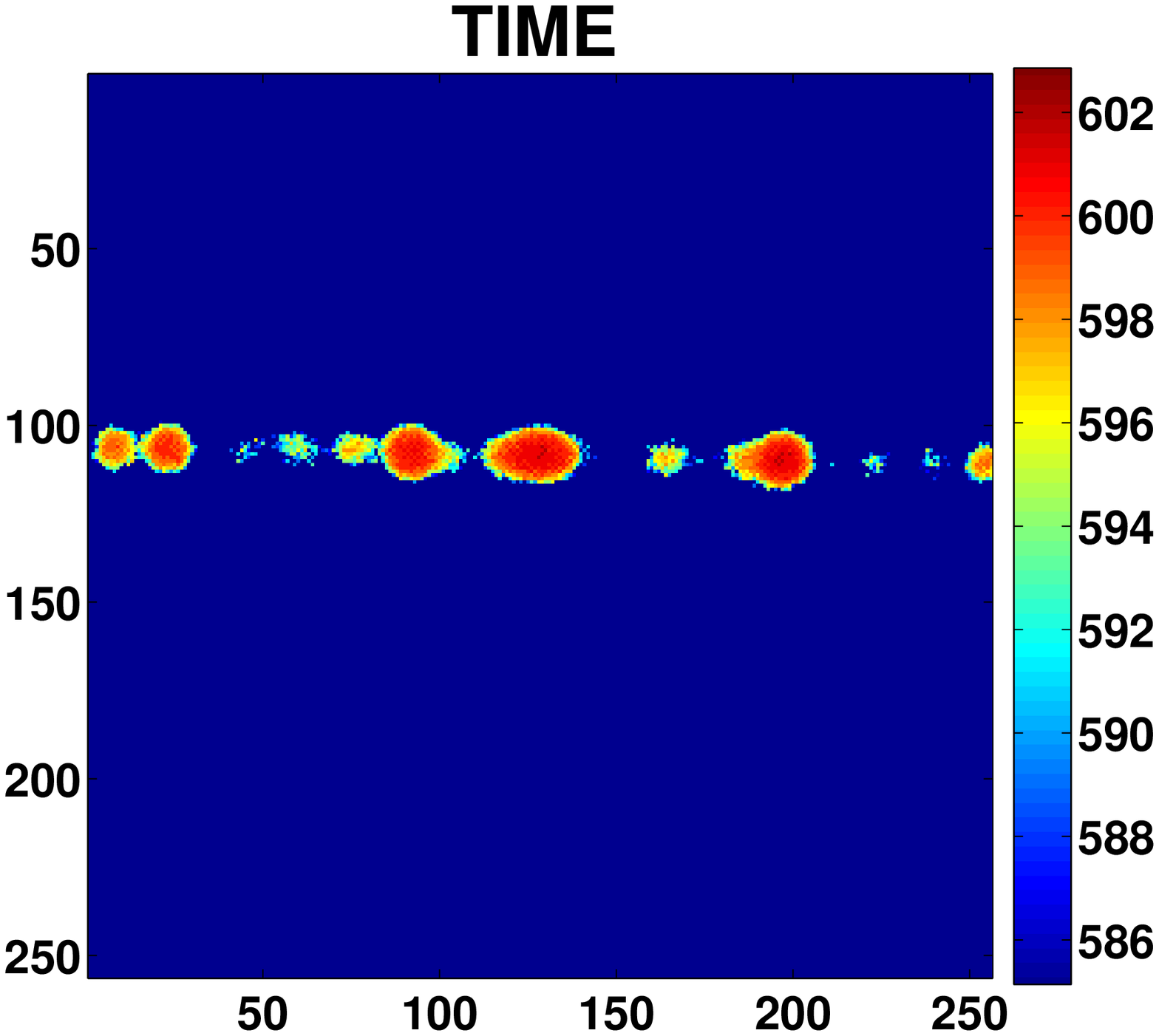}}
 \put(-1.0,52.0){ a) }
 \put(45.0,52.0){ b) }
 \put(90.0,52.0){ c) }
 \end{picture}
\caption{ a) A schematic drawing of the GEM/Medipix2 (Timepix) detector with an 
electron beam crossing the drift volume.
b) and c) Images of 5~GeV electron tracks in $Ar/CO_2$(70:30) recorded
with Medipix2 and Timepix (``TIME'' mode) chips. 
A straight-line fit to the cluster centers is shown.
The $x$, $y$-axes represent the chip's sensitive area, arranged as a square 
matrix of 256 $\times$ 256 pixels of 55 $\times$ 55 $\mu m^2$ size.}
\label{Medipix2_Timepix_track}
\end{figure}

\setlength{\unitlength}{1mm}
\begin{figure}[bth]
 \begin{picture}(51,51)
 \put(0.0,-5.0){\includegraphics{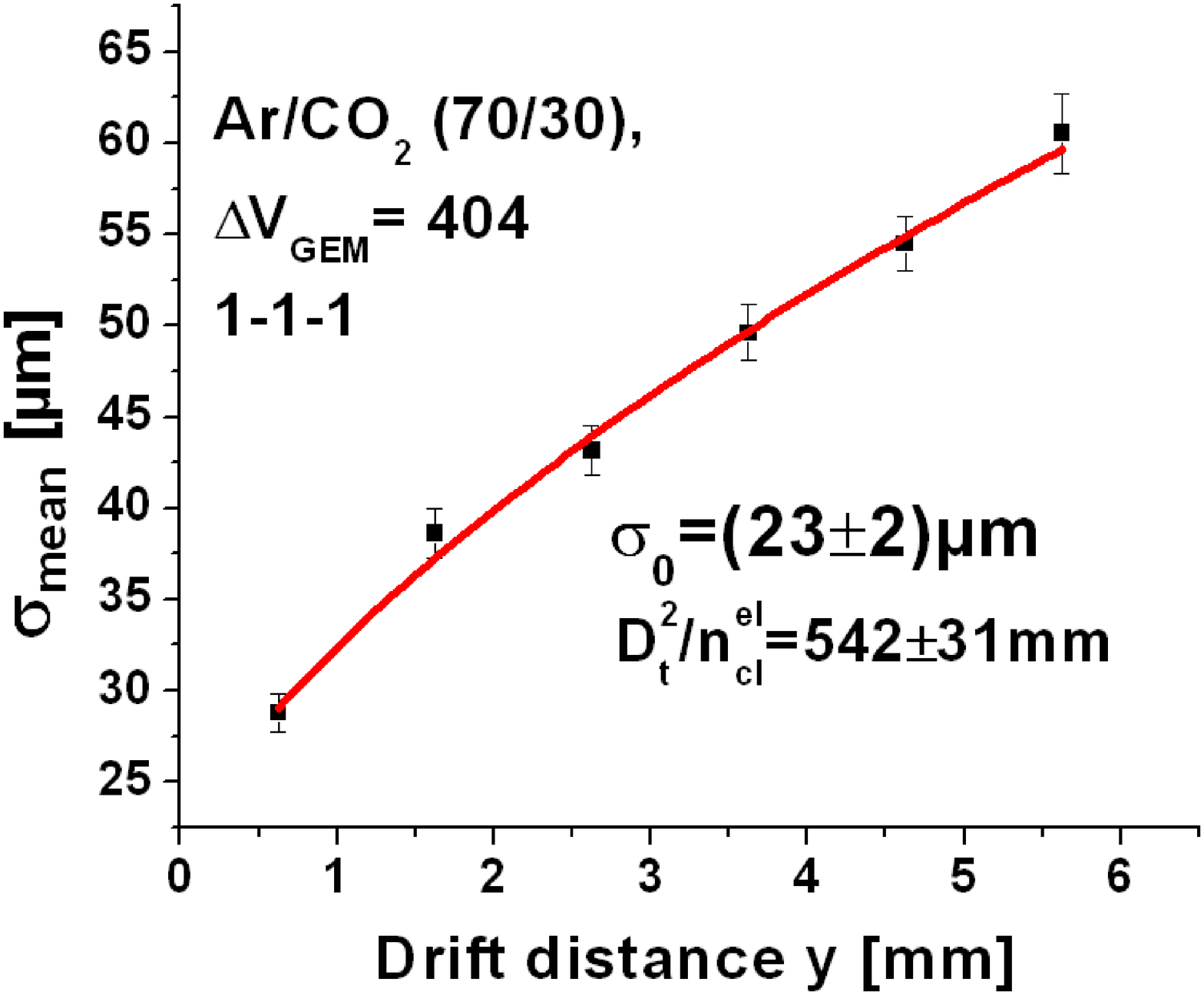}}
 \put(70.0,-3.0){\includegraphics{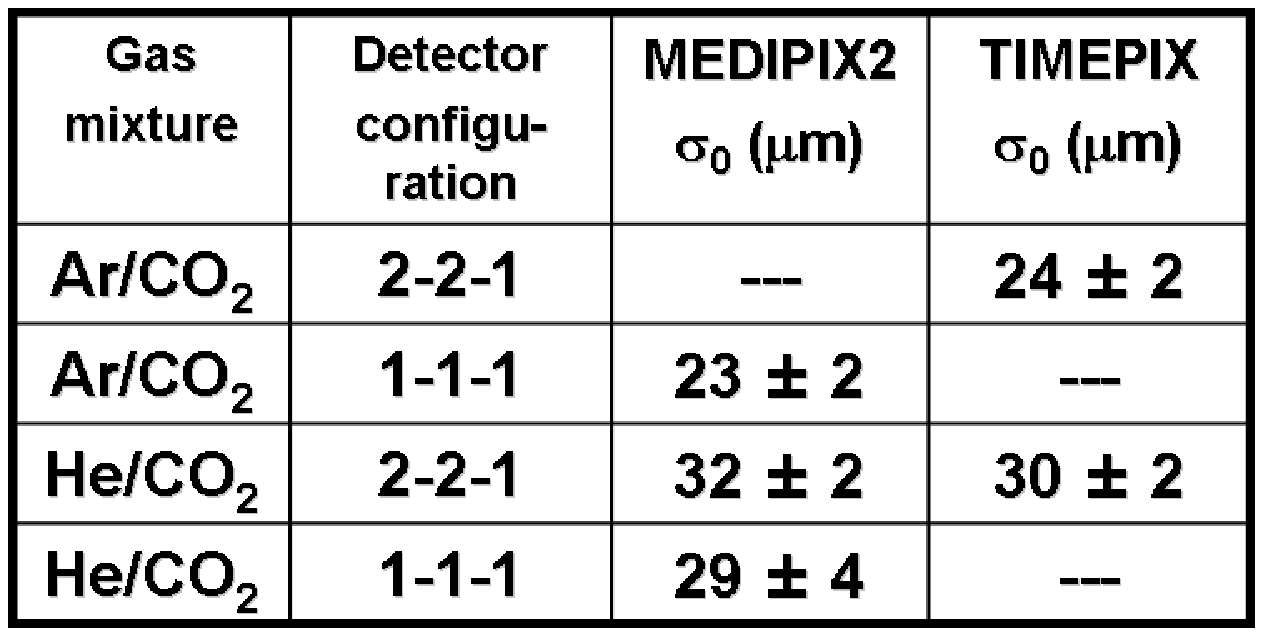}}
 \put(-1.0,45.0){ a) }
 \put(65.0,45.0){ b) }
 \end{picture}
\caption{ a) Spatial resolution ($\sigma_{mean}$) as a function of drift distance 
in an $Ar/CO_2$(70:30) mixture in a triple-GEM/Medipix2 detector.
b) Summary of intrinsic resolution studies ($\sigma_0$) for different 
GEM/Medipix2 and GEM/Timepix stack configurations 
in $Ar(He)/CO_2$(70:30) mixtures.}
\label{summary_medipix2_timepix}
\end{figure}

 Later, the GEM/Medipix2 detector was exposed 
to a 5~GeV electron beam at DESY (see  Fig.~\ref{Medipix2_Timepix_track}a).
 The dependence of the spatial resolution on the drift length 
($y$-coordinate of the track inside the 6~mm GEM drift volume) 
was measured using external $Si$-telescope planes~\cite{bamberger_vienna}.
The deviations of the ``center of gravity'' of individual cluster avalanches
from the straight-line fit can be parameterized as:
\begin{equation}
 \sigma^2_{mean} (y) = \sigma^2_0 + \sigma^2_{dif} (y) = \sigma^2_0 + \frac{D_
t^2 \cdot y}{n_{cl}^{el}},
    \label{digitaltpc}
\end{equation}
where, $\sigma_0$ is the GEM ``defocussing'' term (depends on diffusion of amplified 
electrons in the triple-GEM structure and on the GEM hole/pitch size), 
$D_t$ is a transverse diffusion coefficient,
$n_{cl}^{el}$ is the effective number of primary electrons per cluster
contributing to resolution,
and $y$ is the drift distance from the surface of the first GEM.
 Diffusion of primary ionization clusters as a function of the drift distance 
was clearly observed, as shown in Fig.~\ref{summary_medipix2_timepix}a. 
 The intrinsic spatial resolution, extrapolated to zero drift length ($y$=0),
of $\sigma_0 \approx 23-32~\mu m$ was
measured for different GEM/Medipix2 stack configurations~(see Fig.~\ref{summary_medipix2_timepix}b).
 The triple-GEM device has been also operated with Timepix readout (both
in ``TIME'' and ``TOT'' modes) in the same setup at the DESY beam.
 Fig.~\ref{Medipix2_Timepix_track}c shows an electron track recorded in the
``TIME'' mode; the color denotes the arrival time of electrons in a pixel.
 Similar spatial resolution of $\sigma_0 \approx 24-30~\mu m$
was achieved with the Timepix chip for tracks close to the surface of 
the first GEM (see Fig.~\ref{summary_medipix2_timepix}b). 
 It should also be noted that 
the measured point resolution ($\sigma_0\sim25~\mu m$)
is currently affected by the finite distance between GEM holes ($140\mu m$); 
a better resolution can be achieved with GEMs of smaller pitch.
 These results demonstrate that the CMOS readout of MPGDs meets the
requirements for tracking detectors in the next
generation of high-energy colliders.

\section{Radiation Hardness of Gaseous Detectors}

 Aging phenomena constitute one of the most complex and serious potential problems 
which could limit, or severely impair,
the use of gaseous detectors in unprecedented harsh radiation environments.
 The ``classical aging effects'' are the result of chemical reactions occurring in 
avalanche plasma in wire chambers, which lead to formation of deposits -
conductive or insulating - on the electrode surfaces and manifest themselves 
by a decrease of the gas gain, excessive currents, sparking and 
self-sustained discharges.
   Over the last decade considerable progress has been made in understanding the 
basic rules for the construction and operation of gaseous detectors. 
 Long life in the high-intensity environments of the LHC-era demands not only 
extraordinary radiation hardness of construction materials and gas mixtures but also
very specific and appropriate assembly procedures and quality checks during detector
construction and testing.
Only a limited choice of aging-resistant gases can be successfully used 
at high-luminosity colliders: noble gases, $CF_4$, $CO_2$, $O_2$, $H_2O$. 
Hydrocarbons are not trustable for long-term high-rate experiments.

 A detailed discussion of aging phenomena is beyond the scope of this paper.
  The achievements of past $R\&D$ projects are summarized in~\cite{vavra,kadyk,bouclier,workshop1}, 
%an up-to-date review in the radiation hardness research with state-of-the-art 
%gaseous detectors
the most recent developments in radiation hardness research with state-of-the-art gaseous detectors 
are reviewed in~\cite{workshop2,titov1,titov2}.
Detailed studies of micro-pattern GEM and Micromegas concepts have
revealed that they might be even less vulnerable to radiation-induced performance
degradation in harsh radiation environments than standard silicon microstrip detectors.
 Recently, reliable operation has been established for GEM detectors
coupled to VLSI analog ASICs or Medipix2 CMOS chips; no single pixel chip 
has been destroyed after several months of running~\cite{bellazzini3,freiburg1}.
 More than twenty years of intensive research aimed at matching the needs of high-luminosity
colliders have demonstrated that if properly designed and constructed, 
gaseous detectors can be robust and stable in the presence of high rates and 
heavily ionizing particles. 

\section{Micro-Pattern Gas Detector Applications}

 The performance, robustness and radiation hardness of MPGDs
have encouraged their applications in many other fields; a short 
summary of the most recent results is given below.

\subsection{Gaseous Photomultipliers}

 Large-area RICH gaseous detectors with a thin photosensitive CsI-layer 
deposited on the cathodes of a MWPC are currently employed 
for particle identification in many 
high-energy physics experiments~\cite{piuz,nappi}.
 In recent years there has been considerable progress in the field of 
photon detection by combining MPGDs with semi-transparent 
or reflective CsI photocathodes (PC)
to localize single photoelectrons~\cite{nima502_195}.
 These detectors offer high gain even in noble gases, sub-nanosecond time response, and
excellent localization properties and are able to operate in high magnetic fields and at
cryogenic temperatures.
 Using a triple-GEM detector with hexagonal readout,
 a position accuracy of 55~$\mu m$ and a two-photon separation of around 1~mm
have been achieved~\cite{nima553_18}.
 A Micromegas filled with a $He/iC_4H_{10}$ mixture at atmospheric pressure
allows to achieve a time resolution of $\sim$700~ps for single photoelectrons~\cite{derre_csi}.
 The hole-type gaseous structures: GEMs and Cappillary Plates coupled to CsI-PC
can operate stably down to 80~K~\cite{nima535_517,tns52_927,pavlychenko}.

 The operation of MPGD-based photomultipliers in $CF_4$ with CsI-PC
could form the basis of new-generation windowless Cherenkov detectors
where both the radiator and the photosensor operate in the same gas.
 Exploiting this scheme a Hadron Blind Detector (originally proposed 
for a Parallel Plate Avalanche Chamber~\cite{nima310_585,nima346_120,snic})
has been recently developed and constructed using a triple-GEM amplification system
as part of the upgrade program for the PHENIX experiment at RHIC~\cite{nima546_466,woody}. 
 Hadron blindness is achieved by reversing the direction of the drift field
$E_D$, therefore pushing primary ionization produced by charged particles towards the mesh.
In this configuration photoelectrons released from the CsI surface are still 
efficiently collected into the GEM holes and multiplied.
 The avalanche confinement within the GEM holes 
strongly reduces photon-mediated secondary processes in $CF_4$
(CsI is sensitive to the $CF_4$ scintillation peak at 170~nm)~\cite{nima483_670}.

 The success of GEMs and glass capillary plates 
triggered the development of coarse and more robust structures,
``optimized GEMs''~\cite{nima478_377,tns50_89} followed by 
think-GEM (THGEM)~\cite{nima535_303,nima558_475} gaseous multipliers made of standard PCB
perforated with sub-millimeter diameter holes etched at their rims.
 Effective gas amplification factors of $10^5$ and $10^7$ and fast pulses of a few
nanoseconds rise-time were reached in single and 
cascaded double-THGEM elements.
 Stable operation with high single-photoelectron detection efficiency 
was recorded at fluxes exceeding MHz/mm$^2$.
 A novel spark-protected version of a thick GEM with electrodes made of resistive
kapton (RETGEM) has been recently developed~\cite{oliveira}.
 At low counting rates the detector operates as an conventional THGEM with metallic electrodes
while at high intensities and in case of discharges the behavior is similar to that of a
resistive-plate chamber.
  Recent studies of photosensitive RETGEMs with CsI deposited directly on the dielectric kapton
(without metallic substrate) have shown a rather high 
quantum efficiency (34~$\%$ at 120~nm)~\cite{peskov_vienna}.
 Application of the THGEM and RETGEM concepts to the RICH technology promises to enhance its particle identification capabilities.

\subsection{Optical Readout of GEM-based Detectors}

 Scintillation light emitted during the development of electron avalanches 
can be effectively used for the optical readout of gaseous detectors.
 The luminescence processes in GEMs read out by high-resolution CCDs or by 
PMTs have been used for imaging X-rays, $\alpha$-particles 
and thermal neutrons~\cite{nima513_379}.
 Particular attention has been given to mixtures emitting in the visible
and near-infrared regions (from 400~nm to 1000~nm), the sensitivity region of the CCD.
 One of the most promising candidates having adequate scintillation spectra 
is the $Ar/CF_4$ (95:5) mixture, where $\sim$~0.7 photons above 400~nm are emitted per secondary electron~\cite{nima504_88}.
 The possibility to use a GEM-based scintillation readout of a TPC
as a 3D tracking detector
 has been reported in~\cite{fetal}.
 The double GEM filled with $Ar/CF_4$ (95:5) and 
coupled to a 2$\times$2 matrix of 39-mm diameter PMTs, 
was tested with $^{241}$Am 5.48~MeV $\alpha$-particles, 
emitted under various track angles ($\theta$) with respect to the drift plane.
 The arrival positions of the primary electrons in the GEM plane 
were determined from the center of gravity of the PMT light pulses, analyzed in  
9 equal time intervals as shown in Fig.~\ref{scintillating_GEM}a.
 A typical track angle resolution of $2^\circ$ ($\sigma$) at 
$\theta \sim 45^\circ$ 
and a spatial resolution of better than 1~mm (FWHM) were achieved.

\setlength{\unitlength}{1mm}
\begin{figure}[bth]
 \begin{picture}(51,51) 
\put(2.0,-17.0){\includegraphics{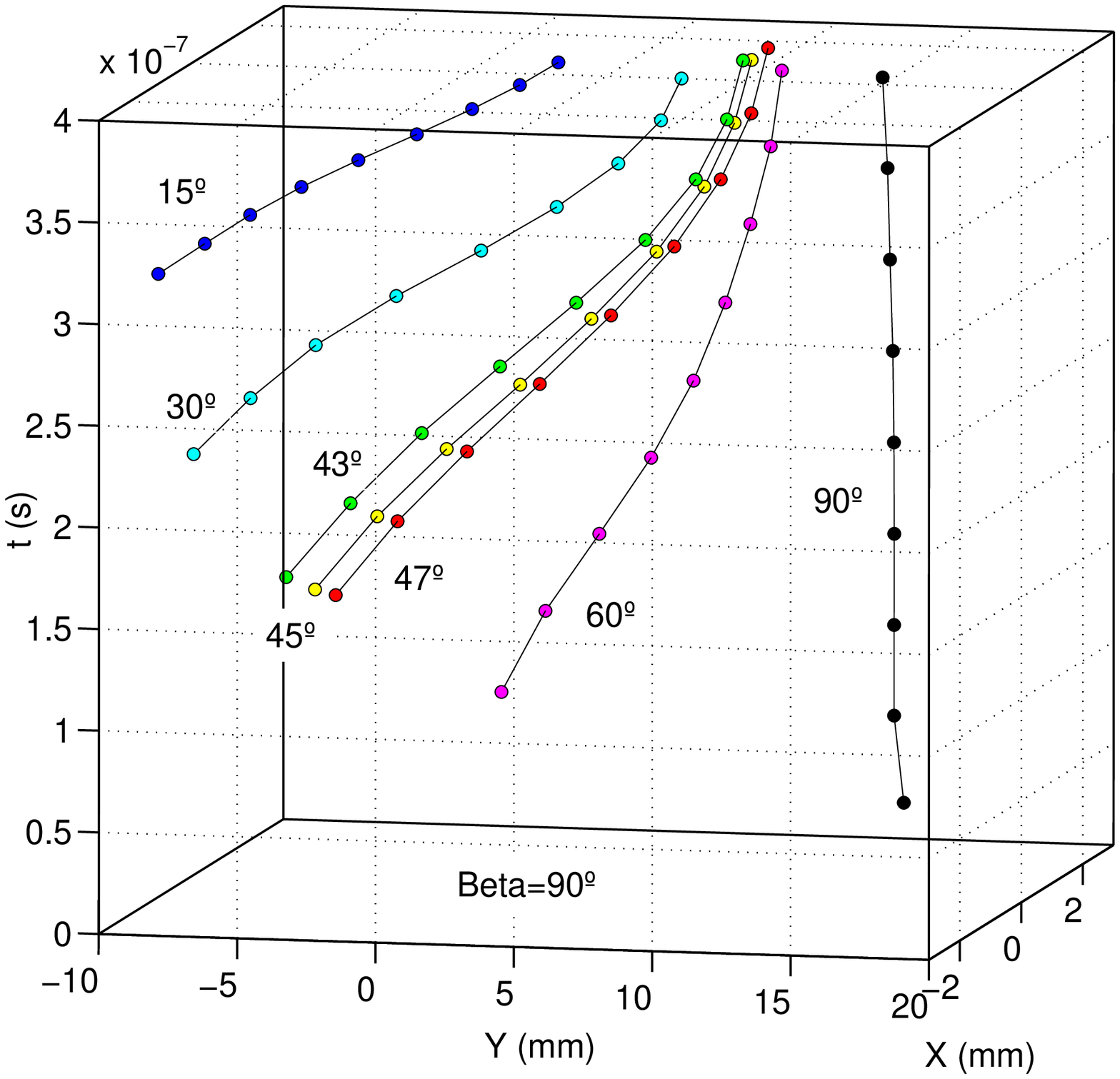}}
\put(72.0,-2.0){\includegraphics{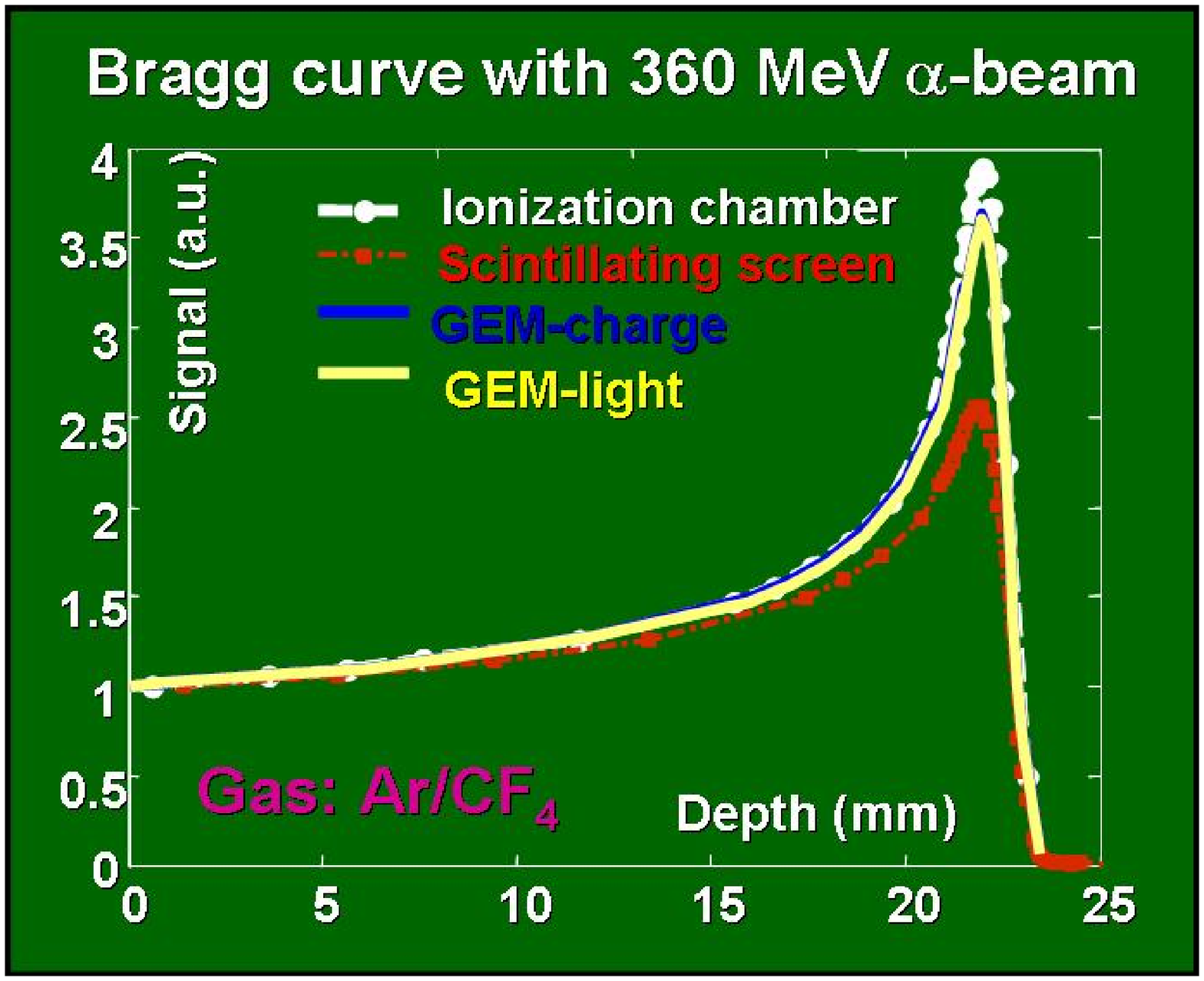}}
 \put(-1.0,50.0){ a) }
 \put(65.0,50.0){ b) }
 \end{picture}
\caption{ a) 3D track reconstruction in the GEM TPC using optical readout.
Individual track PMT pulses were analyzed in 9 time samples. Each 
dot corresponds to the center of gravity of one sample along the track.
b) Comparison between Bragg curves measured with the reference ionization
chamber, the GEM detector (charge and light signals), 
and the Lanex screen from~\cite{Shippers_ieeetns}.}
\label{scintillating_GEM}
\end{figure}

 Another application of the GEM scintillation detector is radiation therapy, which 
demands new online beam-monitoring systems 
with $\sim$~1~mm position resolution and
3-D dosimetry of delivered doses with an accuracy of $\sim$5~$\%$. 
The light yield of a Lanex scintillating screen coupled to a CCD camera, 
which is currently used for the quality control of clinical beams, 
can underestimate the dose by $\sim$30~$\%$ over the last part of
proton track (the Bragg peak)
with respect to the reference ionization chamber.
 A double GEM detector coupled to a CCD camera has been developed to detect photons
emitted by $Ar/CF_4$ (96:4) excited molecules;
the total charge extracted from the GEM holes was also measured~\cite{nima513_42}.
  The intensity of the measured light pattern gives directly a 2-D distribution of the energy deposited 
in the sensitive GEM volume by primary electrons, integrated over all the beam time.
The 3$^{rd}$ dimension can be obtained 
by placing different thicknesses of tissue-equivalent material in front of the detector.
 With a 360~MeV  $\alpha$-beam the integrated light yield was found to be linearly proportional 
to the total charge extracted from the holes of a 2$^{nd}$ GEM, and the scintillating GEM light
signal at the Bragg peak depth
was only 4~$\%$ smaller than that of the reference ionization chamber 
(see Fig.~\ref{scintillating_GEM}b)~\cite{Shippers_ieeetns}.
  Consequently, the scintillating GEM with CCD readout
may become a feasible substitute for the Lanex screen, especially at high 
ionization densities of alpha or carbon-ion beams.

\subsection{Micromegas for Neutron Detection and Low Background Experiments}

 There are many applications of the Micromegas concept in
the neutron detection domain, which include neutron beam diagnostics~\cite{nima524_102}, 
inertial fusion experiments~\cite{nima557_648}, thermal neutron tomography~\cite{andriamonje}
and a novel compact sealed Picollo-Micromegas detector
designed to provide in-core measurements of the neutron flux and 
energy (from thermal to several MeV) in nuclear reactors~\cite{piccollo}.
 Neutrons can be converted into charged particles to detect ionization in Micromegas 
by two means: either using the detector gas filling or a
target with appropriate deposition in its entrance window.
 Micromegas detectors are also used in searching for solar axions (CAST)~\cite{abbon} and
under development for low energy neutrino experiments 
(HELLAZ, NOSTOS)~\cite{gorodetsky,nima530_330}, 
including measurements of neutrino oscillations
and the neutrino magnetic moment.
 In particular, in the CAST experiment at CERN, 
the expected signal comes from solar axion conversions into 
low-energy photons of 1-10~keV energy.
A Micromegas detector with high granularity anode elements
can largely reduce the background event rate down to 5$\times10^{-5}$keV$^{-1}$cm$^{-2}$s$^{-1}$, 
exploiting its stability, good energy and spatial resolution~\cite{ribas}.

\section{Summary and Outlook}

 Today's LHC gaseous detectors have opened a new era of state-of-the-art 
technologies and are the benchmarks for developments beyond the LHC.
 Advances in photo-lithography and micro-processing techniques in the
chip industry during the past decade 
triggered a major transition
in the field of gas detectors 
from wire structures to micro-pattern devices.
 The GEM and Micromegas detectors became a wide-spread tool 
for high-rate tracking over large sensitive areas, 
precision reconstruction of charged particles in the TPC,
$X$-ray, UV and visible photon detection and neutron spectroscopy.
 Modern, sensitive and low-noise electronics will enlarge the range of applications.
As a part of this development, a micro-pattern gas detector
with finely segmented CMOS readout can be used
as an ultra-high-precision ``electronic bubble chamber'' (see Fig.~\ref{timepix_image}), 
opening a window to new physics for many 
applications, especially for low-energy 
charged-particle and photon detection.

\setlength{\unitlength}{1mm}
\begin{figure}[bth]
 \begin{picture}(55,55)
 \put(2.0,-5.0){\includegraphics{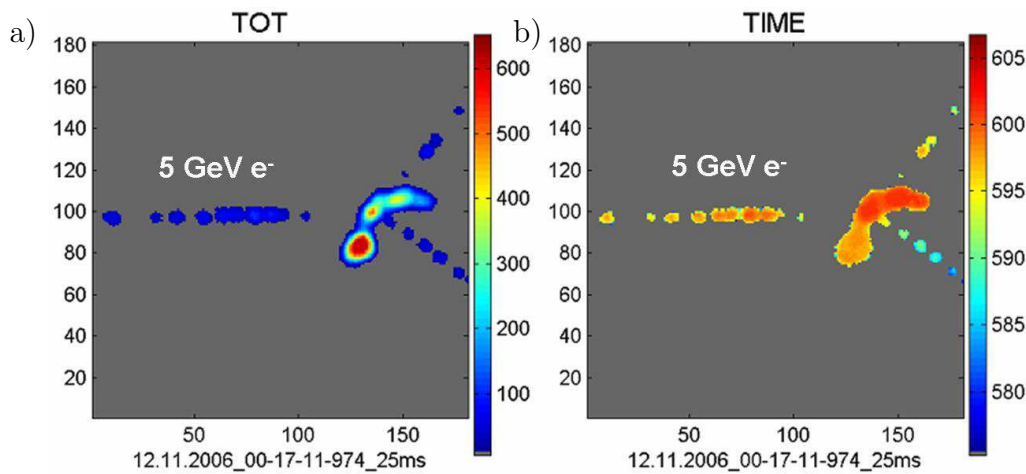}}
 \put(0.0,58.0){ a) }
 \put(67.0,58.0){ b) }
 \end{picture}
\caption{ Inelastic interaction of the 5~GeV electron at the DESY testbeam recorded with
triple-GEM detector readout with a TimePix CMOS chip operated in mixed mode:
every second pixel is operated in the ``TOT'' or ``TIME'' modes, respectively, in a ``chess board'' fashion.
The $x$, $y$-axes represent the chip's sensitive area, obtained by mapping the original data
(matrix of 256 $\times$ 256 pixels of 55~$\mu m$ pitch) onto a 181 $\times$ 181 pixels
matrix with a pitch of 78~$\mu m$.
The color is a measure of time-over-threshold and drift time information, respectively.}
\label{timepix_image}
\end{figure}

\section{Acknoledgements}

 I would like to thank M. Jeitler, M. Krammer, W. Mitaroff and M. Regler for providing
a wonderful cultural and scientific environment during the Vienna Conference.
 I would like to express my gratitude to A. Bamberger, 
R. Bellazzini, P. Colas, K. Desch, I. Giomataris, E. Heijne,
V. Peskov and F. Sauli for many helpful suggestions, stimulating discussions and careful
reading of this manuscript.
 I would like to thank U. Bratzler for reading and correcting this manuscript 
and U. Renz for helping in the preparation of the paper.

\end{document}